\documentclass[aps, superscriptaddress, twocolumn, pre, longbibliography]{revtex4-1}
\usepackage{amsmath, graphicx, color, epsfig, latexsym , bm, ulem, dutchcal , subfigure, float, tikz,eucal, mathpazo, times, braket}
\usepackage[colorlinks=true, linkcolor = blue, urlcolor  = blue, citecolor = blue, anchorcolor = red]{hyperref}   
\usepackage{amssymb}

\usetikzlibrary{positioning}

\begin{document}
	
	
	\title{Quantum thermal rectification via state-dependent two-photon dissipation}
	\author{M. Tahir Naseem} 
    \affiliation{Faculty of Engineering Science, Ghulam Ishaq Khan Institute of Engineering Sciences and Technology, Topi 23640, Khyber Pakhtunkhwa, Pakistan}
	\email{mnaseem16@ku.edu.tr}

	\begin{abstract}{Controlling heat flow at the quantum level is essential for the development of next-generation thermal devices. We investigate thermal rectification in a quantum harmonic oscillator coupled to two thermal baths via both single-photon (linear) and two-photon (nonlinear) exchange processes. At low temperatures, rectification emerges from a state-dependent thermal blockade: the cold bath drives the oscillator into low-occupancy states, suppressing two-photon emission and impeding energy flow. At higher temperatures, rectification is governed by the asymmetric scaling of higher-order moments associated with two-photon absorption and emission. We systematically explore various bath coupling configurations and identify the conditions under which nonlinear dissipation leads to directional heat flow. Furthermore, we propose an implementation scheme based on coupling an auxiliary two-level system to the oscillator, enabling effective two-photon dissipation. We also extend our analysis to three-photon processes and show that rectification increases systematically with photon interaction order. These results contribute to the understanding of quantum heat transport in the presence of nonlinear dissipation and may support future efforts in nanoscale thermal rectification design.}
	\end{abstract}
	
	\maketitle
	
	\section{Introduction}\label{sec:intro}

    The control and understanding of energy flow lie at the heart of both classical and quantum thermodynamics. As technological devices continue to shrink to the nanoscale, thermal effects become increasingly significant, and quantum coherence, fluctuations, and dissipation begin to play dominant roles in the energy transport behavior~\cite{MAHLER200553, Lebon2008, RevModPhys.83.131,Pekola2015}. In such regimes, conventional thermodynamic frameworks must be extended to accommodate the stochastic and quantum nature of microscopic systems~\cite{Vinjanampathy01102016,Goold_2016,Millen_2016}. This shift has motivated intense interest in quantum thermodynamics, where fundamental questions about work, heat, entropy production, and irreversibility are actively being re-examined~\cite{RevModPhys.81.1665, annurev-conmatphys-062910-140506, RevModPhys.83.771, RevModPhys.92.041002}. Concurrently, the ability to manipulate thermal transport at the quantum level is becoming increasingly important for applications ranging from quantum information processing to energy harvesting and nanoscale refrigeration~\cite{RevModPhys.93.041001, Gubaydullin2022, 10.1116/5.0083192, PhysRevLett.108.070604, Aamir2025}. Achieving precise thermal control in these systems is a major challenge, particularly in the presence of unavoidable coupling to the environment. These considerations highlight the need for new strategies to manage heat flow in microscopic devices, including mechanisms to control, suppress, or direct energy transport \cite{Pereira_2019}.

    A variety of strategies have been proposed to regulate heat transport in microscopic systems, including thermal switches, logic gates, transistors, and rectifiers~\cite{10.1063/1.2191730, PhysRevLett.99.177208, RevModPhys.84.1045}. Among them, thermal diodes—which preferentially conduct heat in one direction—have emerged as key components for thermal management in nanoscale devices~\cite{science.1132898, ROBERTS2011648}.
    {\color{black} Numerous theoretical models have been developed to realize thermal diodes, ranging from classical nonlinear lattices~\cite{PhysRevLett.88.094302, PhysRevLett.95.104302, PhysRevE.95.032102} and geometrically asymmetric nanostructures~\cite{science.1132898} to quantum systems.} These include designs based on spin-boson coupling asymmetry~\cite{PhysRevLett.94.034301, PhysRevE.95.022128, PhysRevE.99.042121, PhysRevE.104.054137, PhysRevApplied.15.054050, PhysRevResearch.5.013129}, engineered anharmonic potentials~\cite{PhysRevB.79.144306, PhysRevE.84.061135, Motz_2018, PhysRevResearch.2.033285, PhysRevB.103.155434, PhysRevE.103.012134}, many-body spin systems~\cite{PhysRevE.90.042142, PhysRevE.99.032116, PhysRevE.99.032136, PhysRevE.102.062146}, dynamically modulated quantum circuits~\cite{PhysRevE.99.032126, 10.1063/5.0036485}, and topological structures~\cite{PhysRevApplied.11.044073, Upadhyay_2024}. In the quantum regime, thermal rectification typically relies on breaking reciprocity through geometric or energetic asymmetry, nonlinear dynamics, or a combination thereof \cite{PhysRevE.84.061135}.

    A particularly promising and underexplored mechanism involves nonlinear system-bath interactions, in which the system exchanges multiple quanta with the reservoirs. Such interactions have been widely studied in quantum optics and open quantum systems, where they enable nonclassical state generation, coherence protection, and the engineering of tailored dissipation~\cite{PhysRevA.48.1582, PhysRevLett.79.1467, Eichler2011, PhysRevA.94.033841, PhysRevX.9.021049, PhysRevApplied.22.034053, PhysRevA.110.042411}. However, their role in controlling heat flow and enabling directional energy transport has received comparatively little attention \cite{PhysRevLett.94.034301, 1.1900063, PhysRevB.73.205415, PhysRevE.93.032127}.  Nonlinear dissipation channels can make heat exchange processes intrinsically state-dependent, thereby enabling thermal rectification. Within this broader context, the quantum harmonic oscillator—though incapable of rectifying heat when coupled linearly to thermal baths—provides a powerful and analytically tractable platform to explore these effects when equipped with nonlinear (e.g., two-photon) dissipation mechanisms~\cite{Voje_2013, PhysRevE.96.012114}. This motivates the current investigation, which aims to explore how nonlinear couplings—particularly two-photon dissipation—can be harnessed to achieve tunable quantum thermal rectification. Specifically, we ask: {\it{What are the underlying physical mechanisms that give rise to directional heat flow in systems with two-photon dissipation?}} And {\it{how can such two-photon processes be engineered in experimentally feasible settings?}}

    In this work, we address these questions by analyzing quantum heat transport in a harmonic oscillator coupled to two thermal baths, where each bath can support both single-photon (linear) and two-photon (nonlinear) exchange processes. This model serves as a minimal yet versatile platform for exploring how nonlinearity and dissipation asymmetry affect thermal rectification. A key contribution of this study is the identification of a state-dependent suppression of two-photon emission that leads to asymmetric heat flow under thermal bias reversal. This nonlinearity-induced rectification is particularly prominent in regimes where the cold bath drives the oscillator into low-occupancy states, effectively inhibiting two-photon transitions and thereby creating a thermal bottleneck. At higher temperatures, rectification arises from the nonlinear scaling of higher-order photon number moments. These moments scale differently depending on whether the hot bath is strongly or weakly coupled, resulting in asymmetric energy exchange under temperature reversal. To gain deeper insight into the role of nonlinearity, we extend our analysis to include higher-order photon exchange processes, specifically considering three-photon dissipation. Our results show that thermal rectification increases systematically with the order of the photon exchange. This enhancement arises from stronger thermal blockade effects at low temperatures and increased nonlinear energy flow at higher temperatures.  As a second major contribution, we propose an experimentally feasible scheme to realize two-photon dissipation by coupling an auxiliary two-level system between the harmonic oscillator and the thermal baths. Furthermore, we outline how selective control over single- or two-photon exchange processes can be achieved through reservoir engineering techniques such as bath spectral filtering \cite{PhysRevE.90.022102,Mu_2017,pnas.1805354115,Senior2020,PhysRevResearch.2.033285}. Our findings provide both physical insight and practical routes toward the regulation of thermal flow in quantum systems.

    {\color{black}Thermal rectification has been widely explored when nonlinearity is embedded directly in the working medium, for example, in oscillator lattices with anharmonic potentials~\cite{PhysRevLett.88.094302,PhysRevLett.93.184301,science.1132898}, in superconducting resonators with Kerr-type nonlinearities~\cite{PhysRevB.79.144306,PhysRevApplied.15.054050,PhysRevB.103.155434}, and in optomechanical or trapped-ion platforms with tunable vibrational spectra~\cite{PhysRevE.103.012134,Seif2018}. In these intrinsic-anharmonic schemes, heat-flow asymmetry arises because the transition spectrum depends on occupation or temperature. Left–right asymmetries in coupling or spectral filtering then map into direction-dependent rates upon bias reversal. By contrast, our oscillator remains harmonic and the nonlinearity is introduced at the system--reservoir interface through two-photon dissipation, which yields diode-like transport without modifying the internal Hamiltonian~\cite{Voje_2013,PhysRevE.96.012114}. This delineates a complementary dissipative-engineering route to rectification and complements prior oscillator-based schemes by analyzing a different pathway in quantum systems undergoing nonlinear dissipation.

    Feature by feature, the two architectures differ in mechanism and signatures. Both typically require some left-right asymmetry, for example, unequal couplings or spectral filtering. However, the origin of non-reciprocity is different: intrinsic-anharmonic rectifiers rely on occupation- or temperature-dependent level spacings and detunings, whereas the present scheme relies on two-photon jump operators that conserve parity. At low temperature, Kerr or Duffing devices rectify through activated access to anharmonic transitions, while our model exhibits a two-photon emission blockade when the cold, strongly coupled contact confines the mode to $n<2$. At high temperature, intrinsic-anharmonic rectifiers gain rectification through spectrum shifts and detuning asymmetries, whereas in our model the current is controlled by higher factorial moments and the rectification increases with the order of the multiphoton process. A distinctive signature of the dissipative route is the emergence of parity-sector steady states, which does not occur in purely Kerr rectifiers with linear baths~\cite{Voje_2013,PhysRevE.96.012114}.}

	\begin{figure}[!htbp]
		\begin{center}
			\leavevmode
			\includegraphics[width=0.5\textwidth,angle=0]{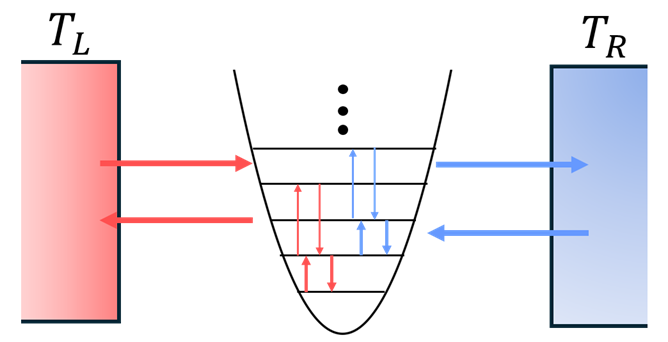}
			\caption{Schematic representation of the model studied in this work. A quantum harmonic oscillator of frequency $\omega$ is coupled to two thermal baths at temperatures \( T_L \) and \( T_R \). Each bath can induce both single-photon and two-photon transitions in the oscillator. The red (left) and blue (right) arrows represent dissipative processes associated with the left and right baths, respectively. Single-photon exchanges allow transitions between adjacent energy levels, while two-photon exchanges involve jumps between levels separated by two quanta.}\label{fig:fig1}
		\end{center}
	\end{figure}

    The remainder of the paper is organized as follows. Sec.~\ref{sec:model} introduces the model and master equation; Sec.~\ref{sec:results} presents steady-state currents and rectification; Sec.~\ref{sec:ExperimentalFeas} outlines a TLS-based implementation; Sec.~\ref{sec:conclusion} concludes. Appendices provide support: Appendix~\ref{app:transients} (transient heat flow and rectification), Appendix~\ref{app:parity} (parity-resolved two-photon steady state), Appendix~\ref{sec:Appendix A} (higher-order photon processes and enhanced rectification),
    Appendix~\ref{app:errorbound} (mean-field-closure error bounds and impact on rectification), and Appendix~\ref{App:B} (microscopic derivation of the master equation and TLS adiabatic elimination).

	\section{The Model}\label{sec:model}

    We consider a single-mode quantum harmonic oscillator of frequency $\omega$ coupled to two distinct thermal baths: a left bath ($L$) and a right bath ($R$). Both baths are modeled as broadband reservoirs that support both single-photon (linear) and two-photon (nonlinear) exchange processes with the system, as shown in Fig. \ref{fig:fig1}. The total Hamiltonian describes the full system
    \begin{equation}
    \hat{H} = \hat{H}_S + \hat{H}_B + \hat{H}_{SB},
    \end{equation}
    where $\hat{H}_S$ is the system Hamiltonian, $\hat{H}_B$ is the Hamiltonian of the two baths, and $\hat{H}_{SB}$ describes the system-bath interactions. The system Hamiltonian is given by
    \begin{equation}
    \hat{H}_S =  \omega \hat{a}^\dagger \hat{a}.
    \end{equation}
    We have set \( \hbar = 1 \) and use this convention in the rest of this work. Here, \( \hat{a} \) and \( \hat{a}^\dagger \) denote the bosonic annihilation and creation operators of the harmonic oscillator mode, satisfying the canonical commutation relation \( [\hat{a}, \hat{a}^\dagger] = 1 \).
    The bath Hamiltonian can be written as
    \begin{equation}
    \hat{H}_B = \sum_{\alpha = L, R} \sum_k  \omega_{\alpha k} \hat{b}_{\alpha k}^\dagger \hat{b}_{\alpha k},
    \end{equation}
    where $\hat{b}_{\alpha k}$ ($\hat{b}_{\alpha k}^\dagger$) are bosonic annihilation (creation) operators for mode $k$ in bath $\alpha \in \{L, R\}$, with frequency $\omega_{\alpha k}$.

    The system-bath interaction Hamiltonian incorporates both one-photon and two-photon exchange terms:
    \begin{equation}
    \hat{H}_{SB} = \sum_{\alpha = L, R} \left[ \hat{a}^\dagger \hat{X}_\alpha^{(1)} + \hat{a} \hat{X}_\alpha^{(1)\dagger} + \hat{a}^{\dagger 2} \hat{X}_\alpha^{(2)} + \hat{a}^2 \hat{X}_\alpha^{(2)\dagger} \right],
    \end{equation}
    where
    \begin{equation}
    \hat{X}_\alpha^{(1)} = \sum_k g_{\alpha k}^{(1)} \hat{b}_{\alpha k}, \quad 
    \hat{X}_\alpha^{(2)} = \sum_k g_{\alpha k}^{(2)} \hat{b}_{\alpha k}.
    \end{equation}
    Here, $g_{\alpha k}^{(1)}$ and $g_{\alpha k}^{(2)}$ are the coupling strengths for the one- and two-photon processes, respectively.

    Under the Born-Markov and secular approximations, and tracing over the bath degrees of freedom, the reduced dynamics of the system can be described by a Gorini–Kossakowski–Sudarshan–Lindblad (GKSL) master equation~\cite{Breuer2002}. {\color{black} We work in the Born–Markov–secular limit; non-Markovian effects may modify heat transport in related settings \cite{10.1063/1.2938092}}. The master equation for the system density matrix $\hat{\rho}$ is given by \cite{Voje_2013}
    \begin{align} \label{eq:fullmaster}
    \frac{d\hat{\rho}}{dt} = -i [\hat{H}_S, \hat{\rho}] + \sum_{\alpha = L, R} \left( \mathcal{L}^{(1)}_\alpha[\hat{\rho}] + \mathcal{L}^{(2)}_\alpha[\hat{\rho}] \right),
    \end{align}
     where $\mathcal{L}^{(1)}_\alpha$ and $\mathcal{L}^{(2)}_\alpha$ are the dissipators corresponding to the one-photon and two-photon processes, respectively:
    \begin{align}
    \mathcal{L}^{(1)}_\alpha[\hat{\rho}] &= \gamma_{\alpha} ( \bar{n}_\alpha + 1 ) \mathcal{D}[\hat{a}]\hat{\rho} + \gamma_{\alpha} \bar{n}_\alpha \mathcal{D}[\hat{a}^\dagger]\hat{\rho}, \\
    \mathcal{L}^{(2)}_\alpha[\hat{\rho}] &= \Gamma_{\alpha} ( \bar{m}_\alpha + 1 ) \mathcal{D}[\hat{a}^2]\hat{\rho} + \Gamma_{\alpha} \bar{m}_\alpha \mathcal{D}[\hat{a}^{\dagger 2}]\hat{\rho}.
    \end{align}
    Here, $\mathcal{D}[\hat{O}]\hat{\rho} = \hat{O} \hat{\rho} \hat{O}^\dagger - \frac{1}{2} \{\hat{O}^\dagger \hat{O}, \hat{\rho}\}$ is the standard Lindblad dissipator. The rates $\gamma_{\alpha}$ and $\Gamma_{\alpha}$ characterize the strength of one- and two-photon interactions with bath $\alpha$, and $\bar{n}_\alpha$ and $\bar{m}_\alpha$ are the thermal occupation numbers at frequencies $\omega$ and $2\omega$, respectively:
    \begin{equation}
    \bar{n}_\alpha = \frac{1}{e^{ \omega / T_\alpha} - 1}, \quad 
    \bar{m}_\alpha = \frac{1}{e^{2 \omega / T_\alpha} - 1},
    \label{eq:nm_def}
    \end{equation}
    with $T_\alpha$ being the temperature of bath $\alpha$ and the Boltzmann constant $k_B $ is set to 1.  {\color{black}Throughout, we distinguish the Bose factors at the one- and two-photon Bohr frequencies:
    $\bar n_\alpha$ and $\bar m_\alpha$. We use $\bar n_\alpha$ exclusively for single-photon channels ($a,a^\dagger$), and $\bar m_\alpha$ exclusively for two-photon channels ($a^2,a^{\dagger 2}$). In addition, the jump $a^2$ couples $|n\rangle \leftrightarrow |n\!\pm\!2\rangle$ and samples bath correlators at the Bohr frequencies $\pm 2\omega$. Accordingly, the upward/downward two-photon rates for bath $\alpha$ read
    \begin{align}
    {\Gamma}_{\downarrow} &\;\propto\; \Gamma_{-}\!\left[\bar n_B(2\omega,T)+1\right],\nonumber\\
     {\Gamma}_{\uparrow} &\;\propto\; \Gamma_{+}\,\bar n_B(2\omega,T),
    \end{align}
    with the KMS detailed-balance relation $G(-\Omega)=e^{-\Omega/T}G(\Omega)$ implying
    \begin{equation}
    \frac{{\Gamma}_{\uparrow}}{{\Gamma}_{\downarrow}}=e^{-2\omega/T}.
    \end{equation}
    Thus, all two-photon terms carry the factor $\bar n_B(2\omega,T)$, denoted $\bar m_\alpha$.

    Our analysis operates in the Born--Markov--secular regime, where flat, memoryless baths yield time-local rates at the Bohr frequencies $\omega$ and $2\omega$. When reservoirs possess finite correlation times or structured spectra, the effective single- and two-photon rates become frequency selective and time-nonlocal, as captured by Nakajima--Zwanzig memory kernels~\cite{10.1143/PTP.20.948,10.1063/1.1731409}. In related models, environmental memory can itself induce directional energy flow and diode-like transients~\cite{Jing2015}, and steady-state heat currents can deviate from GKSL predictions once system--bath correlations contribute to energy storage and exchange~\cite{Chen_2023,RevModPhys.92.041002}. Controlled weak-to-intermediate--coupling treatments beyond standard Redfield (e.g., the modified-Redfield approach)~\cite{PhysRevE.88.052127} and strong-coupling studies with spin baths~\cite{10.1063/1.4871874} show that memory can either enhance or suppress rectification depending on how each contact's spectrum overlaps the system transitions. For the present two-photon mechanism this implies: (i) memory concentrated at the cold contact tends to soften the low-temperature two-photon-emission blockade (repopulating $n\!\ge\!2$ states) and thus reduces rectification; whereas (ii) longer correlation times or enhanced spectral weight near $2\omega$ on the hot side selectively strengthen the nonlinear pathway and increase rectification. These qualitative trends are consistent with broader assessments of quantum thermal machines at strong coupling and beyond the Markov limit~\cite{e18050186}. A fully quantitative analysis of these non-Markovian effects is beyond the scope of this work and is left for future studies.}

	\section{Results} \label{sec:results}

    In this section, we systematically investigate thermal rectification in a quantum harmonic oscillator coupled to two thermal baths, each capable of mediating both single-photon (linear) and two-photon (nonlinear) exchange processes. To isolate and understand the role of different dissipation channels, we consider four distinct cases:

    \begin{enumerate}
    \item \textbf{Only one-photon processes}: We begin by considering the limiting case in which both baths couple to the system solely via single-photon (linear) processes. This case serves as a baseline for comparison and  exhibits no rectification~\cite{PhysRevLett.94.034301}.
    
    \item \textbf{Only two-photon processes}: Next, we examine the scenario where both baths interact with the system exclusively through two-photon (nonlinear) processes. This configuration allows us to highlight the intrinsic asymmetry introduced by multiphoton dissipation channels and their potential to break reciprocity.
    
    \item \textbf{Asymmetric two-photon coupling}: We then analyze a situation in which only one of the two baths supports two-photon exchange processes, while the other is restricted to single-photon coupling. This configuration explicitly breaks the left-right symmetry and is thus expected to exhibit rectification effects.
    
    \item \textbf{Full coupling scenario}: Finally, we consider the most general case in which all four dissipative channels—single- and two-photon processes from both baths—are simultaneously active. This setting offers insight into how linear and nonlinear dissipation jointly influence heat transport and rectification.
    \end{enumerate}

    In each case, we compute the steady-state heat currents under both forward and reverse temperature gradients, and quantify the degree of thermal rectification. The steady-state heat current from bath \( \alpha \in \{L, R\} \) is defined as
    \begin{equation}
    \mathcal{J}_\alpha = \operatorname{Tr} \left[ H_S \mathcal{L}_\alpha(\hat{\rho}_{\text{ss}}) \right],
    \label{eq:HeatCurrentDef}
    \end{equation}
    where \( \mathcal{L}_\alpha \) is the total Liouvillian contribution (including both single- and two-photon processes) from bath \( \alpha \), and \( \hat{\rho}_{\text{ss}} \) is the steady-state density matrix of the system.  
    We adopt the convention that heat current flowing into the system from a bath is considered positive, while heat flowing out of the system into a bath is negative. In the steady state, energy conservation requires \( \mathcal{J}_L + \mathcal{J}_R = 0 \), so we define the net heat current through the system as \( \mathcal{J} = \mathcal{J}_R = -\mathcal{J}_L \), representing the energy inflow from the right bath.

    To quantify thermal rectification, we employ following figure of merit  \cite{PhysRevE.99.042121}
    \begin{equation}\label{eq:FigofMeritRect}
    \mathcal{R} = \frac{|\mathcal{J}_R(T_R, T_L) + \mathcal{J}_R(T_L, T_R)|}{\max(|\mathcal{J}_R(T_R, T_L)|, |\mathcal{J}_R(T_L, T_R)|)},
    \end{equation}
   where \( \mathcal{J}_R(T_R, T_L) \) denotes the steady-state heat current out of the right bath when it is maintained at a higher temperature than the left bath (\( T_R > T_L \)). Conversely, \( \mathcal{J}_R(T_L, T_R) \) corresponds to the heat current when the temperature gradient is reversed (\( T_L > T_R \)). Note that in the absence of rectification, these two currents are equal in magnitude and opposite in sign. The numerator thus measures the degree of asymmetry in the heat flow under reversal of the temperature gradient.
    Consequently, the rectification coefficient \( \mathcal{R} \in [0,1] \), with \( \mathcal{R} = 0 \) indicating no rectification and \( \mathcal{R} = 1 \) corresponding to perfect thermal diode behavior. The numerical simulations in this work are implemented using the QuTiP library~\cite{JOHANSSON20121760}, and the corresponding source code is available at~\cite{CodeRepo}.

\subsection{Only One-Photon Processes} \label{subsec:only1}

    We first consider the case in which both thermal baths are coupled to the system exclusively via one-photon processes, by setting all two-photon coupling rates to zero: \( \Gamma_\alpha = 0 \). This corresponds to a fully linear and symmetric configuration. It is well established from previous studies~\cite{PhysRevLett.94.034301} that such linear systems do not exhibit thermal rectification, and we include this case primarily as a benchmark for comparison with nonlinear scenarios.

    The right bath steady-state heat current is given by~\cite{PhysRevLett.94.034301}
    \begin{equation} 
    \mathcal{J}_R = \operatorname{Tr} \left[ \hat{H}_S \mathcal{L}_R^{(1)}(\hat{\rho}_{\text{ss}}) \right] = \omega \gamma_R \left( \bar{n}_R - n_0 \right),
    \label{eq:J_R1a}
    \end{equation}
    where \( n_0 = \langle \hat{a}^\dagger \hat{a} \rangle \) is the steady-state occupation number of the oscillator and \( \hat{\rho}_{\text{ss}} \) denotes the steady-state density matrix. In the absence of nonlinear dissipation, the occupation number takes the simple form
    \begin{equation}
     n_0  = \frac{\gamma_L \bar{n}_L + \gamma_R \bar{n}_R}{\gamma_L + \gamma_R}.
     \label{eq:n0}
    \end{equation}
    Substituting this into Eq. \eqref{eq:J_R1a}, we obtain the net heat current
    \begin{equation}\label{eq:JR_1}
    \mathcal{J}_R = \omega \frac{\gamma_L \gamma_R}{\gamma_L + \gamma_R} \left( \bar{n}_R - \bar{n}_L \right).
    \end{equation}
    This expression is antisymmetric under temperature exchange \( T_L \leftrightarrow T_R \), implying that the direction of energy flow reverses but its magnitude remains unchanged. As a result, the rectification coefficient \( \mathcal{R} \) vanishes, confirming the absence of thermal rectification in the harmonic systems governed solely by linear (single-photon) dissipation. {\color{black}A complementary analysis of transient heat flow and rectification, including time-to-steady-state estimates, is provided in Appendix~\ref{app:transients}.}

\subsection{Only Two-Photon Processes}\label{subsec:IIIB}

Next, we consider the scenario in which both baths interact with the harmonic oscillator exclusively through two-photon processes, setting the one-photon coupling rates to zero: \( \gamma_\alpha = 0 \). In this regime, the dissipative dynamics are governed entirely by nonlinear interactions. The heat current from the right bath is given by
\begin{align}\label{eq:RHCurrent_2ss}
\mathcal{J}_R &= \operatorname{Tr} \left[ \hat{H}_S \mathcal{L}_R^{(2)}(\hat{\rho}_{\text{ss}}) \right], \nonumber \\
\mathcal{J}_R &= 2 \omega \Gamma_R \Big[ -(\bar{m}_R + 1)\left\langle \hat{n}(\hat{n} - 1) \right\rangle \nonumber \\
&\quad + \bar{m}_R \left\langle (\hat{n} + 1)(\hat{n} + 2) \right\rangle \Big].
\end{align}

    The second-order moments \(\langle \hat{n}(\hat{n} - 1) \rangle\) and \(\langle (\hat{n} + 1)(\hat{n} + 2) \rangle\) arise directly from the two-photon Lindblad jump terms. Emission processes, governed by the operator \(\hat{a}^2\), occur at a rate proportional to \(n(n - 1)\), leading to the appearance of \(\langle \hat{n}(\hat{n} - 1) \rangle\). Conversely, absorption via \((\hat{a}^\dagger)^2\) contributes with a factor \((n + 1)(n + 2)\), yielding the term \(\langle (\hat{n} + 1)(\hat{n} + 2) \rangle\). 
    To compute the heat current, these second-order moments must be evaluated in the steady state, as they determine the net energy exchanged with the baths through two-photon processes.
     Since the oscillator is coupled to both baths via two-photon processes, and the parity of the Fock states is preserved under these dynamics, the steady-state density matrix \(\hat{\rho}_{\text{ss}}\) remains block-diagonal in parity. That is, if the system starts in an even (or odd) parity state, it remains confined to the even (or odd) sector. {\color{black}
    Moreover, because only $n\!\leftrightarrow\!n\pm2$ jumps occur, parity is conserved and $\hat\rho_{\mathrm{ss}}$ is block-diagonal. The even and odd ladders each admit a normalized geometric steady state with the same ratio $r$, but their absolute moments differ (e.g., $\langle \hat n\rangle_{\mathrm{odd}}=\langle \hat n\rangle_{\mathrm{even}}+1$). Consequently, unconditional observables depend on the initial parity weights $w_{\mathrm e},w_{\mathrm o}$. In what follows we initialize the resonator in the even sector ($w_{\mathrm e}=1$), so the two-photon–only closed-form expressions refer to the even ladder. For completeness, Appendix~\ref{app:parity} provides the parity-resolved derivation and gives the odd-sector and mixed-parity steady-state heat currents. When any one-photon channel ($\gamma_\alpha>0$) is present, parity is not conserved and the steady state is unique (sector-independent).}

In the even sector, the relevant Fock states are \(|n\rangle = |2k\rangle\), and we define \(q_k = p_{2k}\) as the probability of being in state \(|2k\rangle\). The two-photon Lindblad terms lead to a second-order recurrence relation for \(q_k\), which at steady state takes the form $q_{k+1} = r q_k$, with a constant ratio
\begin{equation}
    r = \frac{\sum_{\alpha}\Gamma_\alpha \bar{m}_\alpha}{\sum_{\alpha}\Gamma_\alpha (\bar{m}_\alpha + 1)}.
    \label{eq:rdef}
\end{equation}
This recurrence relation arises from \textit{detailed balance} between upward and downward two-photon transitions. At steady state, the rate of absorption-induced transitions \(|2k\rangle \to |2k+2\rangle\) must equal the rate of emission-induced transitions \(|2k+2\rangle \to |2k\rangle\). These rates, derived from the Lindblad jump terms, are proportional to \(\sum_\alpha \Gamma_\alpha \bar{m}_\alpha\) and \(\sum_\alpha \Gamma_\alpha (\bar{m}_\alpha + 1)\), respectively, with identical matrix elements canceling in the ratio. Equating the two determines the constant ratio \(r\) between successive even-parity populations, leading to the geometric distribution \(q_k = (1 - r) r^k\). This steady-state form allows explicit calculation of the average photon number and higher-order moments. The mean occupation number is given by
	\begin{figure}[!htbp]
		\begin{center}
			\leavevmode
			\includegraphics[width=0.49\textwidth,angle=0]{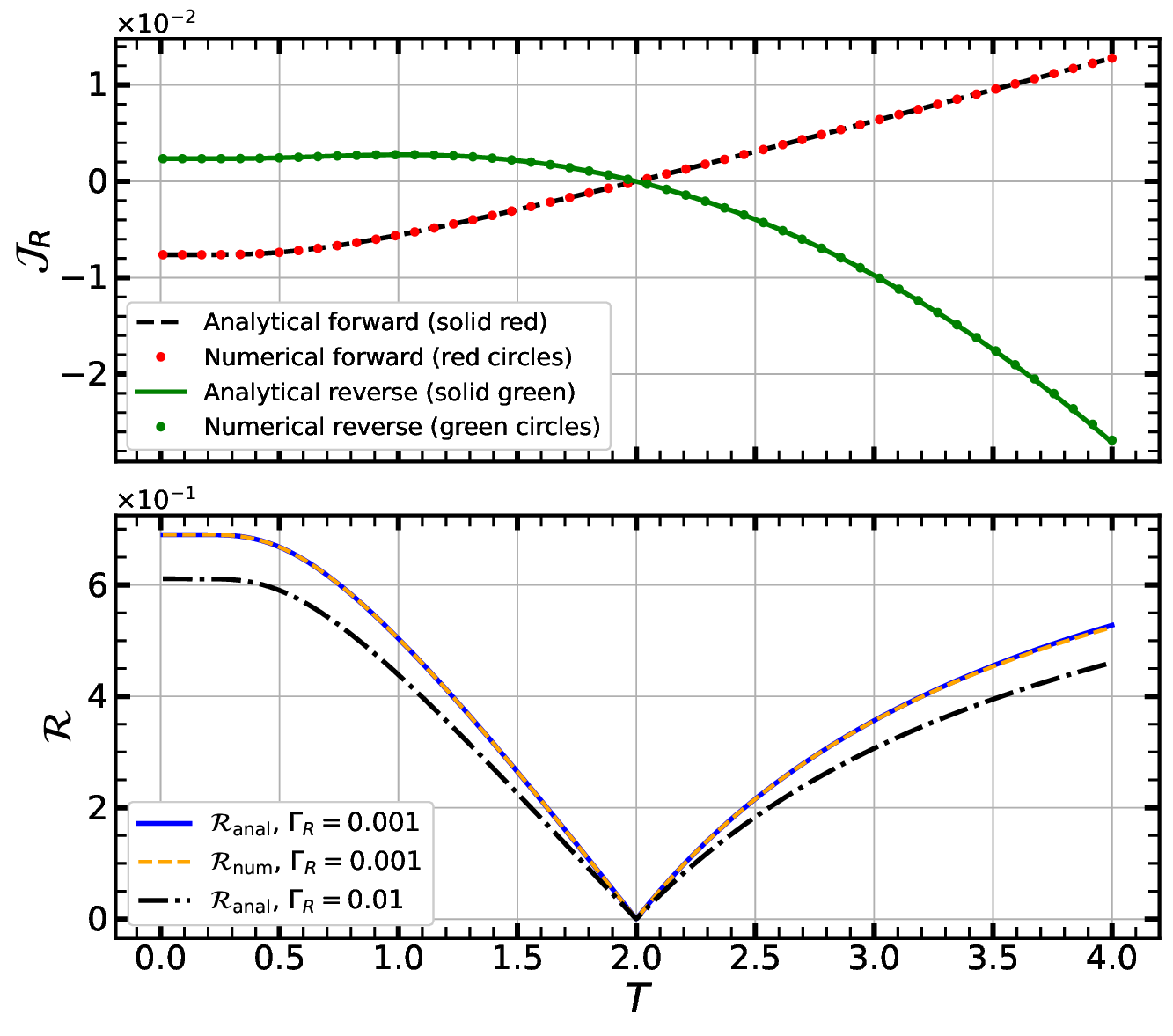}
			\caption{ Heat flow and thermal rectification in the presence of two-photon dissipation. (Top panel) Steady-state heat current \( \mathcal{J}_R \) as a function of temperature \( T \). In the black dashed curve, \( T_L = 2 \) is fixed while \( T_R \) is varied; in the green solid curve, \( T_R = 2 \) is fixed while \( T_L \) is varied.
             The circular markers (red and green) indicate full numerical calculations using the master equation given in Eq. (\ref{eq:fullmaster}) for $\gamma_\alpha = 0$. 
        (Bottom panel) The thermal rectification coefficient \( \mathcal{R} \) as a function of \( T \). 
        The blue solid and orange dashed lines compare analytical and numerical results for \( \Gamma_R = 0.001 \). 
        The black dash-dotted line corresponds to the analytical result for \( \Gamma_R = 0.01 \). 
        Parameters used: \( \omega = 1 \), \( \Gamma_L = 0.1 \).}\label{fig:TwoPhotonRect}
		\end{center}
	\end{figure}
\begin{equation}\label{eq:average}
    \langle \hat{n} \rangle = \sum_{k=0}^{\infty} 2k \, q_k = 2(1 - r) \sum_{k=0}^{\infty} k r^k = \frac{2r}{1 - r}.
\end{equation}
Similarly, we compute the second-order moment:
\begin{align}\label{eq:averages}
\langle \hat{n}^2 \rangle & = \frac{4r(1 + r)}{(1 - r)^2}, \nonumber\\
\left\langle \hat{n}(\hat{n} - 1) \right\rangle &= \frac{2r(1 + 3r)}{(1 - r)^2},\nonumber \\
\left\langle (\hat{n} + 1)(\hat{n} + 2) \right\rangle &= \frac{2(1 + 3r)}{(1 - r)^2}.
\end{align}
{\color{black}Substituting Eqs.~(\ref{eq:average}) and (\ref{eq:averages}) into Eq.~(\ref{eq:RHCurrent_2ss}), we obtain
\begin{align}
\mathcal J_R =4\omega\,\Gamma_R\,\frac{1+3r}{(1-r)^2}\,\Big[\bar m_R(1-r)-r\Big].
\label{R3}
\end{align}
Using the abbreviations
\begin{equation}
A =\Gamma_L\bar m_L+\Gamma_R\bar m_R,\qquad \Sigma_\Gamma=\Gamma_L+\Gamma_R,
\label{R4}
\end{equation}
for which Eq.~(\ref{eq:rdef}) gives \(r=\dfrac{A}{A+\Sigma_\Gamma}\). Then
\begin{equation}
1-r=\frac{\Sigma_\Gamma}{A+\Sigma_\Gamma},\qquad
1+3r=\frac{\Sigma_\Gamma+4A}{A+\Sigma_\Gamma},
\label{R5}
\end{equation}
and
\begin{equation}
\bar m_R(1-r)-r
=\frac{\Sigma_\Gamma\bar m_R-A}{A+\Sigma_\Gamma}.
\label{R6}
\end{equation}
Substituting (\ref{R5})–(\ref{R6}) into (\ref{R3}) and cancelling the common \((A+\Sigma_\Gamma)\) factor,
\begin{equation}
\mathcal J_R
=4\omega\,\Gamma_R\,
\frac{\Sigma_\Gamma+4A}{\Sigma_\Gamma^{\,2}}\,
\big(\Sigma_\Gamma\bar m_R-A\big).
\label{R7}
\end{equation}
Finally, one can write \(\Sigma_\Gamma+4A = \Gamma_L(4\bar m_L+1)+\Gamma_R(4\bar m_R+1)\) and
\(\Sigma_\Gamma\bar m_R-A=\Gamma_L(\bar m_R-\bar m_L)\).} This yields an analytical expression for the steady-state heat current into the right bath:
\begin{align}\label{eq:AnaHeatTwo}
\mathcal{J}_R &= \frac{4 \omega \Gamma_L \Gamma_R}{(\Gamma_L + \Gamma_R)^2} \nonumber \\
&\quad \times \left[ \Gamma_L (4\bar{m}_L + 1) + \Gamma_R (4\bar{m}_R + 1) \right] (\bar{m}_R - \bar{m}_L).
\end{align}

    Unlike the single-photon heat current, described by Eq.~(\ref{eq:JR_1}), which remains antisymmetric under the exchange of bath temperatures regardless of the coupling strengths, the two-photon heat current exhibits a fundamentally different behavior. As shown in Eq.~(\ref{eq:AnaHeatTwo}), it satisfies the antisymmetry relation \( \mathcal{J}_R(T_R, T_L) = -\mathcal{J}_R(T_L, T_R) \) only when the coupling strengths are equal (\( \Gamma_L = \Gamma_R \)). In this symmetric case, forward and reverse currents cancel exactly, and no thermal rectification occurs. However, when the couplings are asymmetric (\( \Gamma_L \ne \Gamma_R \)), this antisymmetry is broken, resulting in different magnitudes of forward and reverse currents and enabling thermal rectification.


    Fig.~\ref{fig:TwoPhotonRect} illustrates the behavior of the steady-state heat current and thermal rectification in the case where only two-photon processes are present. The system is asymmetrically coupled to the baths: the left bath exhibits strong coupling with \( \Gamma_L = 0.1 \), while the right bath is weakly coupled with \( \Gamma_R = 0.001 \) or \( 0.01 \). The top panel displays the heat current \( \mathcal{J}_R \) flowing into the right bath as a function of temperature, whereas the bottom panel shows the corresponding rectification coefficient \( \mathcal{R} \). In contrast to a harmonic oscillator with single-photon interactions with thermal baths—which does not exhibit heat rectification—the present system exhibits qualitatively different behavior. Here, nonlinear two-photon interactions enable directional heat transport, giving rise to thermal rectification. This mechanism can be understood by examining two distinct temperature regimes: low and high temperatures, corresponding respectively to the left and right sides of Fig.~\ref{fig:TwoPhotonRect}.
    
    At low bath temperatures \( T \), the oscillator predominantly occupies low-energy states, resulting in a small average occupation number \( \langle \hat{n} \rangle \). Unlike single-photon processes, two-photon emission requires the oscillator to be in states with at least two excitations, whereas two-photon absorption has no such constraint. When the colder bath is strongly coupled, it tends to drive the oscillator into low-occupancy states with \( n < 2 \), thereby suppressing two-photon emission. This nonlinearity-induced blockade creates a thermal bottleneck: energy absorbed from the hot, weakly coupled bath cannot be efficiently transferred to the cold bath, resulting in a reduced heat current. This suppression is evident on the left side of the solid green curve in the upper panel of Fig.~\ref{fig:TwoPhotonRect}. It corresponds to a configuration with a cold left bath that is strongly coupled (\( \Gamma_L = 0.1, T_L < T_R \)) and a hot right bath that is weakly coupled (\( \Gamma_R = 0.001 \), \( T_R = 2 \)).
    In contrast, the left side of the dashed black curve in the upper panel of Fig.~\ref{fig:TwoPhotonRect} shows a significantly larger heat current. This corresponds to the reversed configuration, where the colder right bath is weakly coupled (\( \Gamma_R = 0.001, T_R< T_L \)) and the hotter left bath is strongly coupled (\( \Gamma_L = 0.1 \), \( T_L = 2 \)). In this case, the weakly coupled cold bath does not efficiently deplete excitations from the oscillator, allowing it to sustain a higher average occupation number \( \langle \hat{n} \rangle \). As a result, both two-photon absorption and emission become more accessible, leading to enhanced thermal transport compared to the case with a strongly coupled cold bath.
    Thus, the asymmetry in emission rates caused by the nonlinear blockade leads to thermal rectification in the low-temperature regime.

    At high temperatures \( T \), both two-photon emission and absorption become significant, scaling as \( \langle \hat{n}(\hat{n} - 1)\rangle \) and \( \langle (\hat{n} + 1)(\hat{n} + 2)\rangle \), respectively. The heat current is particularly enhanced when the hotter bath is strongly coupled, as seen on the right side of the solid green curve in the upper panel of Fig.~\ref{fig:TwoPhotonRect}, compared to the weaker coupling case shown by the right side of the dashed black curve. 
    
    These results show that thermal rectification in the two-photon coupling case arises from the combined effects of (i) asymmetric system-bath coupling and (ii) the intrinsic nonlinearity of two-photon processes.
    Notably, asymmetry in coupling alone is insufficient to induce rectification, as demonstrated in systems with only single-photon interactions. The essential distinction lies in the fact that, in the nonlinear regime, the energy transfer rates depend on higher-order moments of the occupation number, which scale differently with temperature and coupling strength. At low temperatures, rectification arises due to the suppression of two-photon emission, while at high temperatures, rectification results from the enhancement of nonlinear response, particularly when the hotter bath is strongly coupled. To further explore this mechanism, we extend our analysis to include three-photon dissipation processes in Appendix~\ref{sec:Appendix A}, and demonstrate that rectification becomes even more pronounced with increasing photon exchange order.

  {\color{black}
    With the closed-form steady-state moments in hand (Eqs.~\eqref{eq:average}–\eqref{eq:averages}), we convert the qualitative bottleneck picture into a quantitative crossover criterion. We then translate this criterion into an experimentally accessible cold-side temperature using the definition of $r$ (Eq.~\eqref{eq:rdef}). Note that, there is no sharp phase transition; rather, the bottleneck lifts through a crossover once the cold-side two-photon emission weight is no longer rate-limiting. To make this operational, we set the two-photon emission factorial moment to unity (cf. Eq.~\eqref{eq:averages}):
    \begin{align}
    \langle \hat n(\hat n{-}1)\rangle = 1.
    \label{eq:threshold}
    \end{align}
    Using the closed-form moments in Eqs.~\eqref{eq:average}–\eqref{eq:averages}, this condition yields
    \begin{align}
    r_c = 0.2, 
    \qquad 
    \langle \hat n\rangle_c = \frac{2 r_c}{1-r_c} = \frac{1}{2}.
    \end{align}
    To translate this crossover into a temperature, we combine the definition of $r$ (Eq.~\eqref{eq:rdef}) with a cold and hot notations for the two baths:
    \begin{align}
    r \;=\; 
    \frac{\Gamma_{\mathrm{cold}}\,\bar m_{\mathrm{cold}}+\Gamma_{\mathrm{hot}}\,\bar m_{\mathrm{hot}}}{
      \Gamma_{\mathrm{cold}}(\bar m_{\mathrm{cold}}+1)+\Gamma_{\mathrm{hot}}(\bar m_{\mathrm{hot}}+1)}.
    \end{align}
    Setting $r=r_c$ and solving for the cold-side Bose factor gives
    \begin{align}
    \bar m_{\mathrm{cold}}
    = \frac{r_c\,(\Gamma_{\mathrm{cold}}+\Gamma_{\mathrm{hot}})}{(1-r_c)\,\Gamma_{\mathrm{cold}}}
      - \frac{\Gamma_{\mathrm{hot}}}{\Gamma_{\mathrm{cold}}}\,\bar m_{\mathrm{hot}},
      \label{eq:mcold}
    \end{align}
    and, using Eq.~\eqref{eq:nm_def}, the corresponding temperature is
    \begin{align}
    T_{\mathrm{cold}}
    =\frac{2\omega}{\ln\!\bigl(1+1/\bar m_{\mathrm{cold}}\bigr)}.
    \label{eq:Tcold}
    \end{align}
    In the limit $\Gamma_{\mathrm{cold}}\!\gg\!\Gamma_{\mathrm{hot}}$, using Eq. \eqref{eq:Tcold}, Eq. \eqref{eq:mcold} simplifies  to
    \begin{align}
    T_{\mathrm{cold}}^{\ast}\simeq \frac{2\omega}{\ln(1/r_c)}=\frac{2\omega}{\ln 5},
    \end{align}
    providing a convenient practical scale for the crossover (e.g., for $\omega=1$, $T_{\mathrm{cold}}^{\ast}\!\approx\!1.24$). The precise numerical value depends only weakly on the $\mathcal{O}(1)$ threshold in Eq.~\eqref{eq:threshold}, so alternative, nearby choices lead to modest shifts in $T_{\mathrm{cold}}^{\ast}$.
    }

    \subsection{Asymmetric two-photon coupling}\label{subsec:IIIc}

    We now examine a hybrid configuration in which the left bath is coupled to the harmonic oscillator via both single- and two-photon exchange processes, while the right bath is coupled exclusively through single-photon interactions. This arrangement introduces asymmetry not only in the coupling strengths but also in the nature of the dissipation mechanisms, enabling the exploration of the interplay between linear and nonlinear processes.

    In this setup, the master equation~(\ref{eq:fullmaster}) is modified by setting \( \Gamma_R = 0 \), so that the right bath interacts with the system solely through one-photon processes. The steady-state heat current flowing into the right bath is then given by  
    \begin{equation}\label{eq:heatsubC}
    \mathcal{J}_R = \text{Tr}\left[\hat{H}_S \mathcal{L}_R^{(1)}(\hat{\rho}_{\text{ss}})\right] = \omega\, \gamma_R \left(\bar{n}_R - \langle \hat{n} \rangle\right),
    \end{equation}  
    where \( \langle \hat{n} \rangle \) is the steady-state average photon number of the oscillator. The left bath’s combination of linear and nonlinear dissipation channels governs the occupation of the oscillator and thereby indirectly controls the heat flow into the right bath. This leads to qualitatively different behavior compared to the case discussed in Subsec.~\ref{subsec:only1}, where the left bath interacts with the oscillator solely via single-photon process.

    To compute the steady-state heat current into the right bath, we evaluate the average photon number \( \langle \hat{n} \rangle \). Its time evolution is obtained by tracing the master equation over the system Hamiltonian and dissipators. In the present configuration, only the left bath supports two-photon transitions, while both baths contribute single-photon processes. The resulting rate equation is  
    \begin{align}
    \frac{d}{dt} \langle \hat{n} \rangle &= 
    \sum_{\alpha = L, R} \gamma_\alpha 
    \left( \bar{n}_\alpha - \langle \hat{n} \rangle \right) 
    + 2 \Gamma_L \big[
    \bar{m}_L \langle (\hat{n} + 1)(\hat{n} + 2) \rangle  \nonumber \\
    &\quad  - (\bar{m}_L + 1) \langle \hat{n}(\hat{n} - 1) \rangle
    \big].
    \label{eq:dn_dt}
    \end{align}
    Due to the presence of two-photon dissipation, the differential equation for the average photon number involves not only \( \langle \hat{n} \rangle \) but also higher-order moments. More generally, the equation for the \( n \)th moment depends on the \( (n+1) \)th moment, leading to an infinite hierarchy of coupled equations~\cite{PhysRev.156.286}. As a result, each moment is tied to both lower and higher-order moments, making the system analytically intractable without further approximations. Specifically, the rate equation for \( \langle \hat{n} \rangle \) in Eq.~\eqref{eq:dn_dt} includes second-order quantities such as \( \langle \hat{n}(\hat{n} - 1) \rangle \) and \( \langle (\hat{n} + 1)(\hat{n} + 2) \rangle \), which obstruct the closure of the equation.

    To proceed analytically, we consider the steady-state regime by setting \( \frac{d}{dt} \langle \hat{n} \rangle = 0 \), and invoke the semi-classical approximation. Under this approximation, the second-order moments are factorized as
    \begin{align}
    \langle \hat{n}(\hat{n} - 1) \rangle &\approx \langle \hat{n} \rangle^2 -   \langle \hat{n} \rangle, \nonumber \\
    \langle (\hat{n} + 1)(\hat{n} + 2) \rangle &\approx \langle \hat{n} \rangle^2 + 3\langle \hat{n} \rangle + 2.
    \label{eq:factor}
    \end{align}
    Substituting these expressions into Eq.~\eqref{eq:dn_dt} yields a closed quadratic equation for \( \langle \hat{n} \rangle \), given by
    \begin{equation}
    \tilde{A} \langle \hat n\rangle ^2 + \tilde{B} \langle \hat n \rangle + \tilde{C} = 0,
    \end{equation}
    with coefficients
    \begin{align*}
    \tilde{A} &= -2\Gamma_L, \quad
    \tilde{B} = -\gamma + 2\Gamma_L (4\bar{m}_L + 1), \nonumber \\
    \tilde{C} &=  \gamma_L \bar{n}_L + \gamma_R \bar{n}_R + 4\Gamma_L \bar{m}_L,
    \end{align*}
    here, $\gamma = \gamma_L +\gamma_R$. For the parameters considered, the physically relevant solution corresponds to the positive root of the quadratic equation. Substituting the analytical expression for the steady-state photon number \( n \) into the expression for the right-bath heat current equation (\ref{eq:heatsubC}) yields a closed-form result:
\begin{equation}
\mathcal{J}_R = \omega \gamma_R \left( \bar{n}_R - \frac{-\tilde{B} - \sqrt{\tilde{B}^2 - 4\tilde{A}\tilde{C}}}{2\tilde{A}} \right).
\label{eq:heatsubC2}
\end{equation}

    The semi-classical approximation is valid when the two-photon dissipation rate \( \Gamma_L \) is much smaller than the single-photon rates, i.e., \( \Gamma_L \ll \gamma_\alpha \). In this regime, the photon number distribution remains approximately thermal, and the steady-state density matrix becomes diagonal in the Fock basis due to dominant one-photon gain and loss processes~\cite{Voje_2013}. However, the approximation's validity also depends on the temperature: at high temperatures, higher-order corrections—particularly higher moments—become significant, and the semi-classical approximation breaks down.
     
    The steady-state heat current \( \mathcal{J}_R \) given in Eq. (\ref{eq:heatsubC2}) exhibits a clear asymmetry under the exchange of bath temperatures \( T_L \) and \( T_R \), indicating the presence of thermal rectification. This asymmetry originates from the unequal nature of system-bath couplings: the left bath supports both single- and two-photon dissipation channels, while the right bath is limited to single-photon processes. Consequently, the heat current expression, which depends on the steady-state photon number \( n \), does not remain invariant under temperature reversal. As a result, swapping \( T_L \) and \( T_R \) does not lead to equal and opposite energy flow, in contrast to systems with fully linear interactions.
	\begin{figure}[!htbp]
		\begin{center}
			\leavevmode
			\includegraphics[width=0.48\textwidth,angle=0]{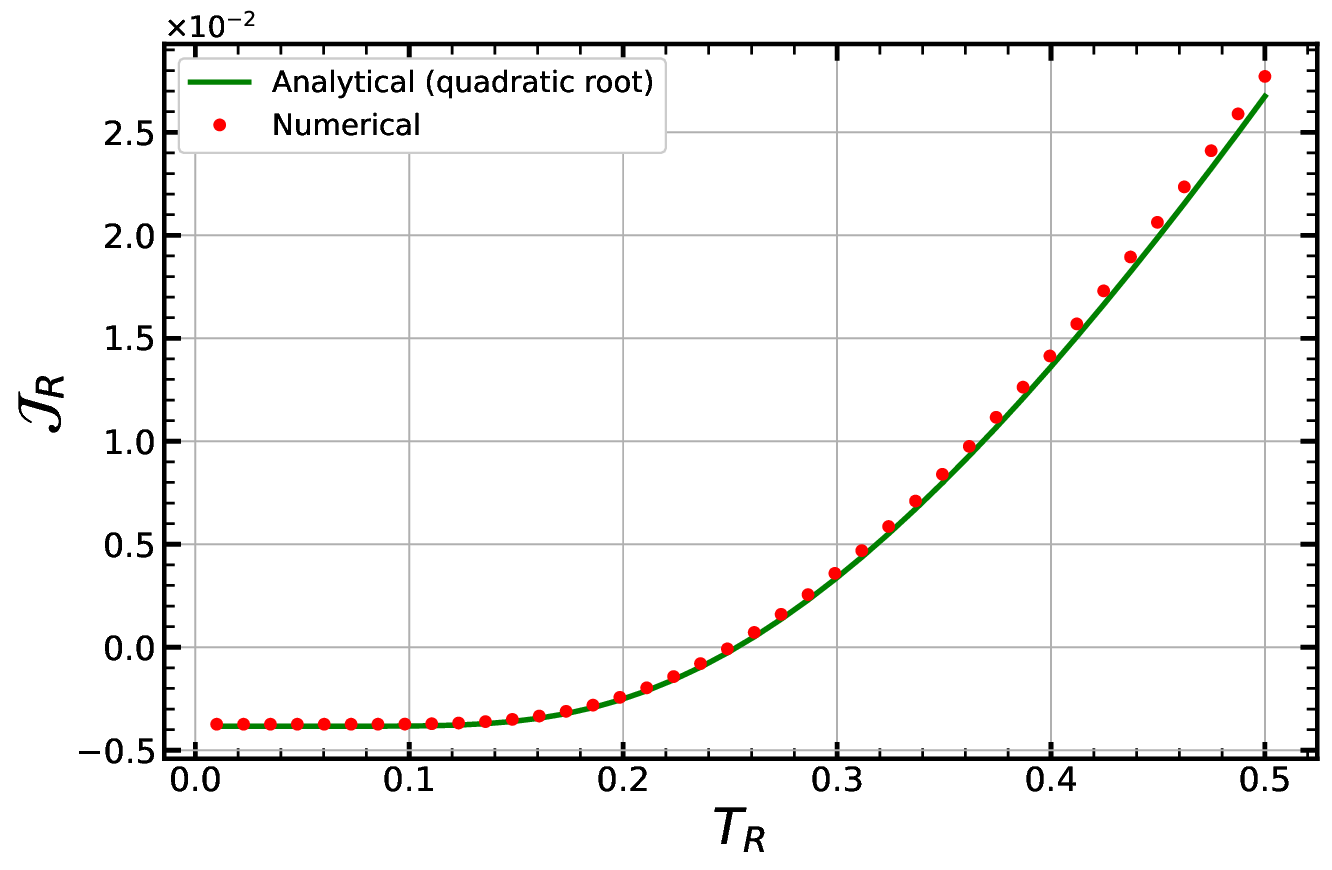}
			\caption{Steady-state heat current $\mathcal{J}_R$ flowing from the right bath, plotted as a function of its temperature $T_R$, with the left bath temperature held fixed at $T_L = 0.25$. The figure compares numerical results (circles) with analytical results (solid green line).
            Parameters: $\omega = 1$, $\gamma_R = 0.4$, $\gamma_L = 0.4$, $\Gamma_L = 0.01$, and Fock space dimension = 50.}\label{fig:fig3}
		\end{center}
	\end{figure}
    When the left bath is hotter (\( T_L > T_R \)), the presence of both dissipation mechanisms enables more efficient excitation of the oscillator. The linear contribution scales with \( \bar{n}_L \), while the nonlinear channel becomes increasingly dominant at higher temperatures due to its dependence on higher-order photon number moments. Conversely, when the right bath is hotter (\( T_R > T_L \)), energy is transferred only through the single-photon channel, which lacks the enhanced thermal response of the nonlinear pathway. This imbalance results in an asymmetric steady-state population \( n  \), and consequently, a direction-dependent heat current \( \mathcal{J}_R \).

   This asymmetry can be more clearly understood by analyzing the limit of very weak two-photon coupling, i.e., \( \Gamma_{L} \ll \gamma_\alpha \). {\color{black} Starting from Eq.~(\ref{eq:dn_dt}) and inserting the factorisations from Eq.~(\ref{eq:factor}), the steady-state condition \(d\langle \hat n\rangle/dt=0\) yields
\begin{align}
0
&=\sum_{\alpha=L,R}\gamma_\alpha\big(\bar n_\alpha- \langle \hat n \rangle\big) \nonumber\\
&\quad +2\Gamma_L\!\Big(\bar m_L\big(\langle \hat n \rangle^2+3\langle \hat n \rangle+2\big)-(\bar m_L+1)\big(\langle \hat n \rangle^2-\langle \hat n \rangle\big)\Big).
\label{eq:dn/dt=0_a}
\end{align}
Invoking the definition of \(n_0\) (Eq.~\eqref{eq:n0}), the steady-state Eq. \eqref{eq:dn/dt=0_a} reduces to a quadratic equation for \(\langle \hat n\rangle\).
\begin{align}
0=\gamma\big(n_0-\langle \hat n \rangle\big)+2\Gamma_L\Big(-\,\langle \hat n \rangle^2+(4\bar m_L+1)\,\langle \hat n \rangle+2\bar m_L\Big).
\label{eq:seriesGammaL}
\end{align}
Assuming very weak nonlinear coupling, we keep only the first-order term in the series expansion \(\langle \hat n \rangle=n_0+\Gamma_L n_1+\mathcal O(\Gamma_L^2)\). Substituting this ansatz into Eq.~\eqref{eq:seriesGammaL} and retaining the first non-vanishing order gives
\begin{align}
n_1=\frac{2}{\gamma}\Big(-\,n_0^{\,2}+(4\bar m_L+1)\,n_0+2\bar m_L\Big).
\end{align}
Using the right-bath heat current, Eq.~\eqref{eq:heatsubC}, and expanding to first order in \(\Gamma_L\), we obtain
\begin{align}
\mathcal{J}_R
&=\omega\,\frac{\gamma_L\gamma_R}{\gamma}\,(\bar n_R-\bar n_L)\nonumber\\
&\quad +\omega\,\frac{\Gamma_L\gamma_R}{\gamma}\Big(2n_0^{\,2}-2(4\bar m_L+1)n_0-4\bar m_L\Big)
+\mathcal O(\Gamma_L^2).
\label{eq:heatsubC2_approx}
\end{align}
In this expression, only \(\bar m_L\) appears in the first-order correction because, with \(\Gamma_R=0\), both the upward and downward two-photon contributions originate solely from the left reservoir. This makes the \(T\)-reversal asymmetry explicit already at order \(\mathcal O(\Gamma_L)\).
}  In the Eq. \eqref{eq:heatsubC2_approx}, the first term corresponds to the conventional heat current through a harmonic oscillator coupled to two thermal baths via single-photon exchange, as given in Eq.~ (\ref{eq:JR_1}). This term is symmetric under temperature exchange \( T_L \leftrightarrow T_R \) and does not contribute to rectification, as the direction of heat flow simply reverses under thermal bias inversion.
    The second term, arising from the two-photon interaction with the left bath, introduces asymmetry. While the \( n_0 \) component of this correction is symmetric under \( T_L \leftrightarrow T_R \) when \( \gamma_L = \gamma_R \), the term involving \( \bar{m}_L \) breaks this symmetry. Since \( \bar{m}_L \) depends only on the temperature of the left bath, and the right bath does not support two-photon coupling, there is no corresponding \( \bar{m}_R \) term to balance the expression. This imbalance results in a heat current that is asymmetric under temperature exchange.
    Therefore, the presence of two-photon coupling exclusively in the left bath is essential for the emergence of thermal rectification. In the absence of such coupling (\( \Gamma_L = 0 \)), the correction term vanishes, and the heat current becomes fully symmetric, regardless of the values of \( \gamma_L \) and \( \gamma_R \). This analysis demonstrates that rectification in this setup arises from the interplay between nonlinearity and asymmetry in the dissipation mechanisms.

    Fig.~\ref{fig:fig3} shows the steady-state heat current \( \mathcal{J}_R \) flowing into the right bath as a function of its temperature \( T_R \), with the left bath held fixed at \( T_L = 0.25 \). Numerical results (circles) are compared with analytical predictions from Eq.~(\ref{eq:heatsubC2}). As expected, the heat current vanishes at \( T_R = T_L \) and increases with thermal bias. The analytical expression agrees well with the numerical results in the weak two-photon coupling regime, but deviations emerge at higher \( T_R \), where nonlinear effects become more pronounced. {\color{black}Appendix~\ref{app:errorbound} benchmarks the mean-field factorization used in the hybrid asymmetric case and provides analytic error bounds on its effect on the heat current and rectification. 
}

\begin{figure}[!htbp]
		\begin{center}
			\leavevmode
			\includegraphics[width=0.48\textwidth,angle=0]{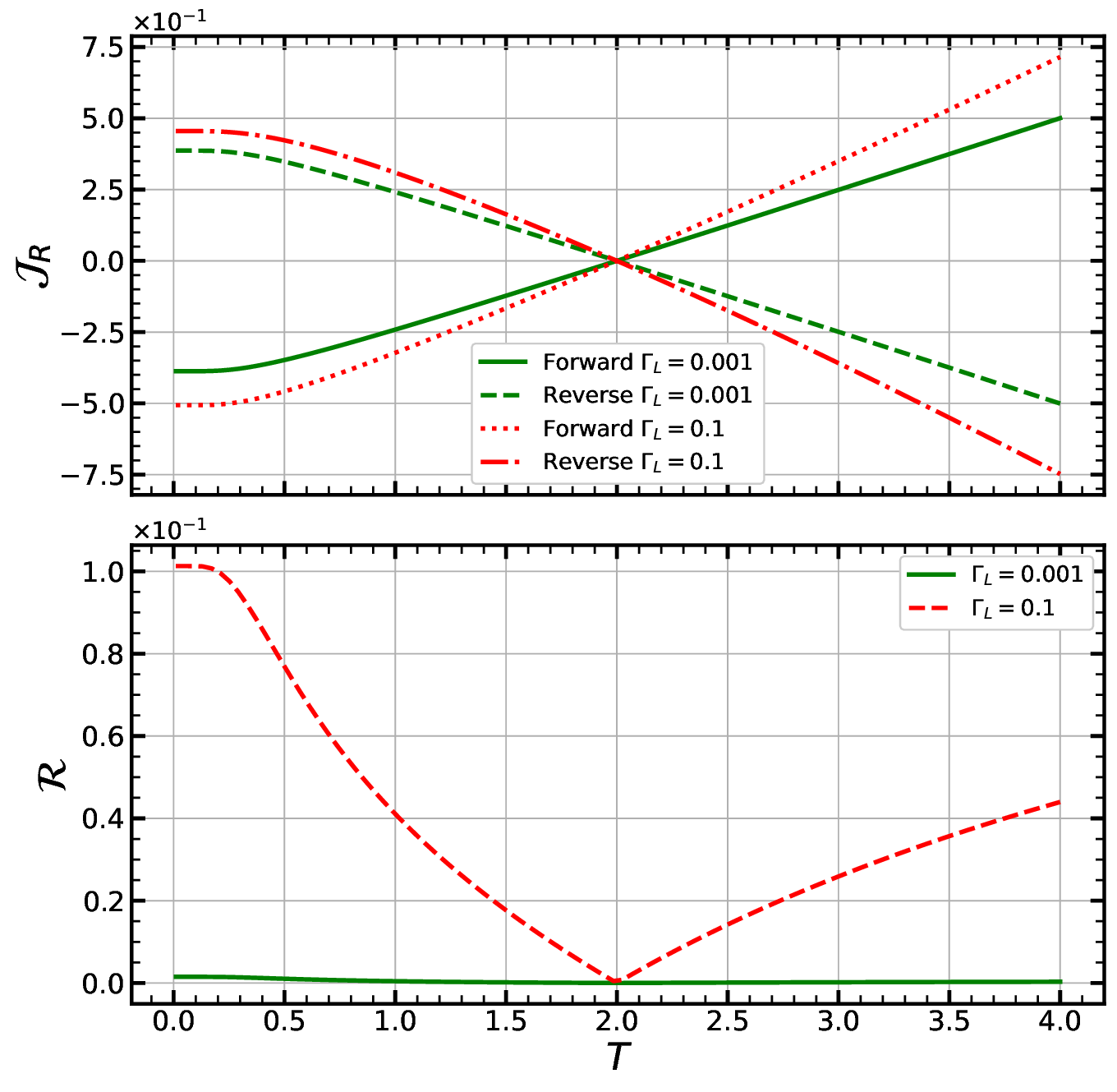}
			\caption{Heat current and rectification for $\Gamma_R = 0$. The forward configuration corresponds to fixed left bath temperature $T_L = 2$ and varying $T_R$, while the reverse configuration interchanges the temperatures. 
            Top: Steady-state heat current $\mathcal{J}_R$ flowing into the right bath as a function of $T_R$, evaluated for two values of the two-photon dissipation rate: $\Gamma_L = 0.001$ (solid and dashed green curves) and $\Gamma_L = 0.1$ (dotted and dash-dotted red curves). 
            Solid and dotted lines represent the forward configuration, while dashed and dash-dotted lines represent the reverse configuration. 
             Bottom: Corresponding rectification coefficient quantifying the asymmetry of heat transport under temperature exchange. 
             System parameters: oscillator frequency $\omega = 1.0$, and one-photon dissipation rates $\gamma_L = \gamma_R = 0.5$.}\label{fig:fig4}
		\end{center}
	\end{figure}

    Fig.~\ref{fig:fig4} presents the steady-state heat current \( \mathcal{J}_R \) and the corresponding rectification coefficient for \( \Gamma_R = 0 \). In the top panel, two configurations are compared: the forward direction (solid and dotted lines), where \( T_L = 2 \) is fixed and \( T_R \) is varied, and the reverse direction (dashed and dash-dotted lines), where the temperatures are interchanged. {\color{black}When the two-photon dissipation strength is very weak (\( \Gamma_L = 0.001 \); green curves), the forward and reverse currents almost coincide in absolute value (magnitude) and differ only by sign, indicating negligible thermal rectification.} This occurs because transport is dominated by the single-photon channel, which is symmetric under temperature exchange.
    
    However, when the two-photon coupling becomes stronger (\( \Gamma_L = 0.1 \); red curves), a noticeable asymmetry develops between the forward and reverse heat currents, resulting in a measurable rectification coefficient, as shown in the bottom panel of Fig.~\ref{fig:fig4}. This indicates that nonlinear interactions begin to influence thermal transport significantly.
    Despite this, the observed rectification remains weaker than in the case where only two-photon processes are active and single-photon dissipation is absent (see Fig.~\ref{fig:TwoPhotonRect}). The key difference lies in the system's behavior at low temperatures. As previously discussed, strong rectification arises when two-photon emission from the colder bath is suppressed due to insufficient thermal excitation, since such processes require at least two quanta. This leads to an effective blockade of reverse heat flow.
    In contrast, when single-photon processes are also present, energy can still be exchanged with the cold bath through one-photon transitions, even when the oscillator remains in a low-excitation state. As a result, the asymmetry that drives rectification is diminished: while the two-photon channel introduces directionality, the one-photon pathway provides a symmetric ``bypass'' that allows thermal transport in both directions.

    \subsection{Full coupling scenario}
    
    We now turn to the most general case, in which both thermal baths are coupled to the harmonic oscillator through a combination of single-photon (linear) and two-photon (nonlinear) dissipation processes. This configuration incorporates all four system-bath interaction channels, enabling  investigation of how linear and nonlinear mechanisms jointly influence steady-state heat transport and thermal rectification.

    The steady-state heat current flowing into the right bath in this configuration is given by
    \begin{align}
    \mathcal{J}_R &= \text{Tr}\Big[\hat{H}_S \sum_{i=1,2} \mathcal{L}_R^{(i)}   (\hat{\rho}_{\text{ss}})\Big] \nonumber \\
    &= \omega\, \gamma_R \left(\bar{n}_R - \langle \hat{n} \rangle\right) 
    + 2 \omega \Gamma_R \big[ -(\bar{m}_R + 1)\left\langle \hat{n}(\hat{n} - 1) \right\rangle \nonumber \\
    &\quad + \bar{m}_R \left\langle (\hat{n} + 1)(\hat{n} + 2) \right\rangle \big].
    \end{align}

\begin{figure}[!htbp]
		\begin{center}
			\leavevmode
			\includegraphics[width=0.48\textwidth,angle=0]{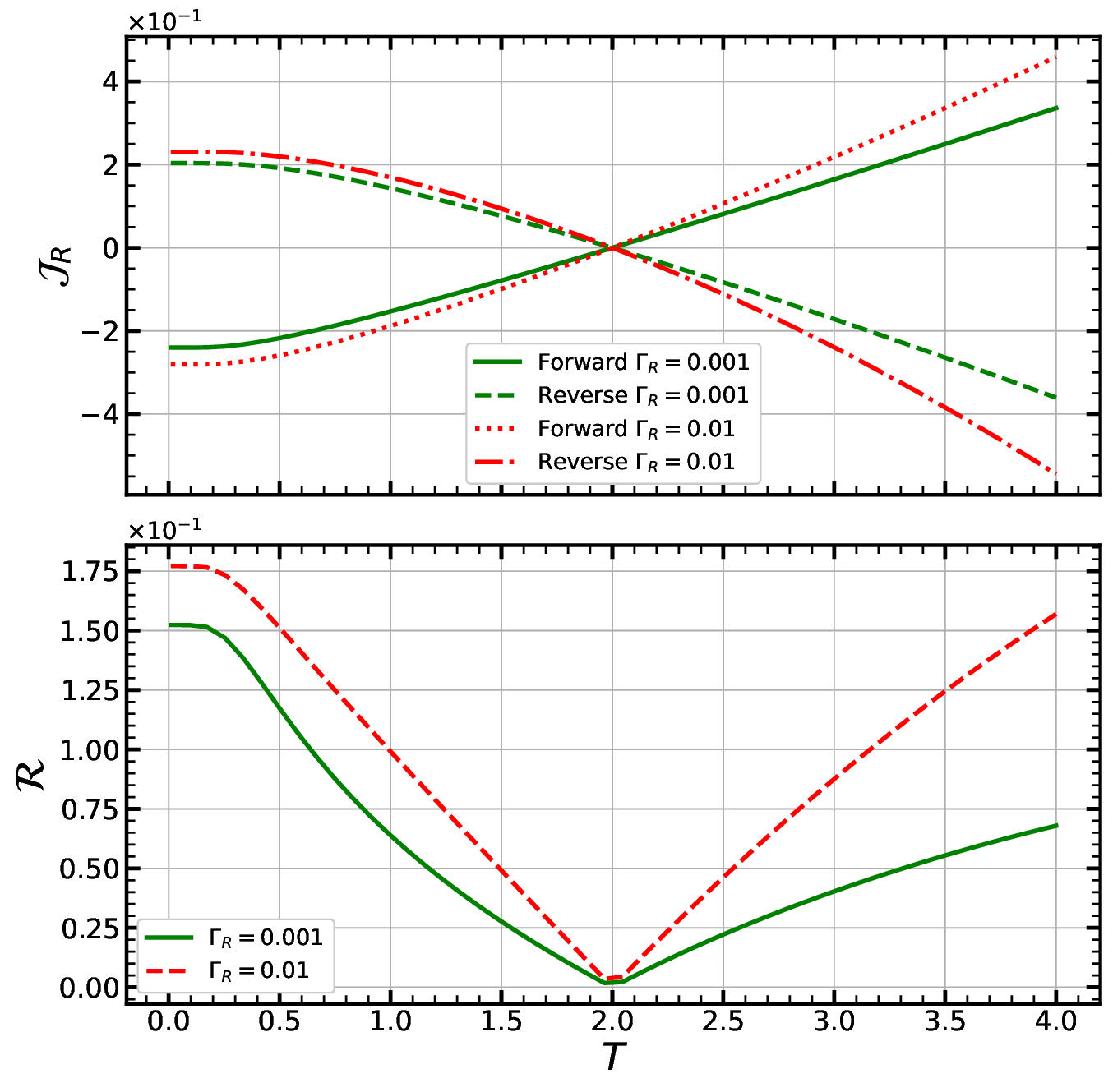}
			\caption{ Steady-state heat current \(\mathcal{J}_R\) (top) and corresponding rectification coefficient \(\mathcal{R}\) (bottom) as a function of temperature \(T\). The forward bias configuration corresponds to the case where the left bath is fixed (\(T_L = 2.0\)) while \(T_R\) is varied; in the reverse bias configuration, \(T_R = 2.0\) is fixed and \(T_L\) is varied. The left bath is coupled to the harmonic oscillator via both single- and two-photon processes, with \(\gamma_L = 0.2\) and \(\Gamma_L = 0.1\), while the right bath supports both dissipation channels with \(\gamma_R = 0.2\) and varying two-photon coupling strength \(\Gamma_R = 0.001, \text{and}\, 0.01\). Solid and dotted lines show forward bias configurations, while dashed and dash-dotted lines represent the reverse configuration. Increasing \(\Gamma_R\) enhances the nonlinear response of the right bath, leading to a pronounced asymmetry in the heat current under temperature exchange and resulting in higher rectification.}\label{fig:fig5}
		\end{center}
	\end{figure}

    In the presence of both dissipation mechanisms at each contact, we compute the steady-state heat current and thermal rectification numerically by solving the full master equation~(\ref{eq:fullmaster}). The results are shown in Fig.~\ref{fig:fig5}, where the right-bath heat current \( \mathcal{J}_R \) and rectification coefficient \( \mathcal{R} \) are plotted as functions of bath temperature. In the forward bias configuration, the left bath temperature is fixed at \( T_L = 2.0 \), and \( T_R \) is varied. In the reverse configuration, \( T_R = 2.0 \) is held fixed while \( T_L \) is varied.

    Compared to the purely two-photon case presented earlier in Fig.~\ref{fig:TwoPhotonRect}, the heat current here remains antisymmetric under temperature exchange. However, its magnitude is less sensitive to nonlinear effects due to the presence of linear dissipation pathways in both baths. These channels facilitate energy transport even when nonlinear processes are suppressed—particularly at low temperatures where the two-photon emission blockade would otherwise limit current. As a result, the degree of rectification observed in Fig.~\ref{fig:fig5} is generally weaker than that in the purely nonlinear case.
    Nonetheless, rectification becomes more pronounced as \( \Gamma_R \) is increased, owing to the enhanced two-photon coupling strengths. This behavior is consistent with the mechanism illustrated in Fig.~\ref{fig:TwoPhotonRect}, where rectification emerges from the nonlinear nature of multiphoton interactions. Unlike the configuration in Fig.~\ref{fig:fig4}, where only one bath supports two-photon processes and rectification arises from explicit asymmetry in dissipation types, the present case involves symmetric coupling structures on both sides. Here, rectification results from a more subtle interplay between thermal bias, nonlinear dissipation, and the balancing effect of linear processes.

	\section{Implementing nonlinear dissipation via an auxiliary system}\label{sec:ExperimentalFeas}
	
 The nonlinear coupling between a harmonic oscillator and its environment—responsible for two-photon dissipation—can be engineered using an auxiliary system. Here, we outline a scheme to realize an effective nonlinear system-environment interaction based on an auxiliary two-level system (TLS). Specifically, we consider a TLS coupled to a harmonic oscillator (HO) via an energy-field interaction~\cite{PhysRevB.79.041302, PhysRevLett.115.203601}, described by the Hamiltonian
\begin{equation}\label{eq:tlsho}
\hat{H}_\text{s} = \frac{\omega_{0}}{2}\hat{\sigma}_{z} + \omega_{a}\hat{a}^{\dagger}\hat{a} + g\hat{\sigma}_{z}(\hat{a} + \hat{a}^{\dagger}),
\end{equation}
where \( \omega_{0} \) and \( \omega_{a} \) are the frequencies of the TLS and HO, respectively, and \( g \) is the coupling strength. The creation and annihilation operators are \( \hat{a}^\dagger \) and \( \hat{a} \), and \( \hat{\sigma}_{k} \) denotes the Pauli matrices.
We assume the TLS is coupled to a thermal bath, indirectly connecting the HO to this environment. The bath Hamiltonian is
\begin{equation}
\hat{H}_\text{b} = \sum_{k}\omega_{k}\hat{b}^{\dagger}_{k}\hat{b}_{k},
\end{equation}
and its interaction with the TLS has the form
\begin{equation}\label{eq:sb}
\hat{H}_\text{sb} = \sum_{k}g_{k}\hat{\sigma}_{x}(\hat{b}_k + \hat{b}_k^{\dagger}),
\end{equation}
where \( g_k \) is the coupling between the TLS and the \( k \)-th bath mode.
    
{\color{black}
To isolate the reduced dynamics of the HO, we first derive a Lindblad-form master equation for the joint HO–TLS system under the Born–Markov and secular approximations, and then adiabatically eliminate the TLS. The resulting HO-only master equation, presented in Appendix~\ref{App:B}, reads
\begin{align}
\frac{d\tilde{\rho}_a}{dt}
&= \tilde{\gamma}_-\, D[\tilde{a}] 
 + \tilde{\gamma}_+\, D[\tilde{a}^\dagger] 
 + \tilde{\Gamma}_-\, D[\tilde{a}^2] 
 + \tilde{\Gamma}_+\, D[\tilde{a}^{\dagger 2}]\nonumber\\
  &+ \tilde{\gamma}_d\, D[\tilde{a}^\dagger \tilde{a}],
\end{align}
where we neglect the coherent Lamb-shift contribution. The effective rates  \(\tilde{\gamma}_\pm\), \(\tilde{\Gamma}_\pm\), and \(\tilde{\gamma}_d\) are provided in Appendix~\ref{App:B}.}
This master equation represents the desired form for realizing engineered dissipation that supports both single- and two-photon processes. The first two terms, \( \tilde{\gamma}_- D[\tilde{a}] \) and \( \tilde{\gamma}_+ D[\tilde{a}^\dagger] \), correspond to conventional single-photon emission and absorption, respectively. The terms \( \tilde{\Gamma}_- D[\tilde{a}^2] \) and \( \tilde{\Gamma}_+ D[\tilde{a}^{\dagger 2}] \) describe nonlinear two-photon emission and absorption processes, capturing the essential physics required for studying heat transport mediated by multiphoton interactions.
The final term, \( \tilde{\gamma}_d D[\tilde{a}^\dagger \tilde{a}] \), represents pure dephasing in the Fock basis. While this term can influence the system’s coherence properties, it does not contribute to energy exchange between the system and the environment. Consequently, it does not affect the steady-state heat current and can be neglected for the purposes of our study. Dropping this term leaves us with a master equation that accurately captures both single- and two-photon dissipation channels relevant to steady-state thermal transport.

In addition to enabling two-photon dissipation, our scheme also offers the flexibility for selective control over energy exchange processes. By tuning system parameters or exploiting the spectral properties of the environment, specific dissipation channels can be enhanced or suppressed. This can be achieved through \textit{reservoir engineering} techniques such as spectral filtering~\cite{PhysRevE.90.022102, Mu_2017, pnas.1805354115, Senior2020, PhysRevResearch.2.033285, PhysRevE.90.022102, PhysRevResearch.2.033285, Naseem_2020, Naseem2021, Naseem_2022, PhysRevA.105.012201}. For instance, coupling the system to a narrow-band resonator centered at the frequency \( \omega_{-2} = \omega_0 - 2\omega_a \) effectively filters out single-photon sidebands while allowing two-photon transitions to persist~\cite{Naseem2021, Wang2024}. This configuration realizes a dissipation model dominated by two-photon processes, as investigated in Subsec.~\ref{subsec:only1}. Conversely, coupling to a resonator tuned to \( \omega_{-} = \omega_o - \omega_a \) suppresses two-photon exchanges while retaining single-photon transitions~\cite{PhysRevA.105.012201, Naseem_2022}. These spectral filtering methods thus provide a versatile and experimentally feasible route to access different dissipative regimes and enable controlled implementation of the theoretical models explored in this work.

	\section{Conclusions}\label{sec:conclusion}

    We have investigated quantum thermal rectification in a harmonic oscillator coupled to two thermal baths through both single- and two-photon exchange processes. By systematically exploring various configurations—including purely linear, purely nonlinear, and hybrid dissipation scenarios—we have demonstrated that thermal rectification arises from the interplay between nonlinearity and asymmetry in system-bath couplings. In particular, we identified a state-dependent suppression mechanism that dominates at low temperatures: when the colder bath is strongly coupled, it confines the oscillator to low-excitation states, thereby suppressing two-photon emission and creating a thermal bottleneck. At higher temperatures, rectification is driven by the asymmetric scaling of higher-order moments associated with two-photon processes, resulting in direction-dependent heat flow. Extending this framework, we have shown in Appendix~\ref{sec:Appendix A} that rectification becomes increasingly pronounced as the order of the photon exchange process increases, with three-photon interactions producing an even stronger rectification effect.

    Our results show that even weak nonlinearities can induce measurable rectification when combined with asymmetry in bath coupling strengths. To support experimental relevance, we have proposed an implementation scheme in which effective two-photon dissipation is realized by coupling the oscillator to an auxiliary two-level system connected to thermal reservoirs. Additionally, we have discussed how reservoir engineering techniques—such as bath spectral filtering—can be used to selectively suppress single- or two-photon channels, enabling experimental realization of the scenarios studied here. While our analysis focuses on a minimal model, the physical mechanisms uncovered are broadly applicable to a wide class of quantum systems exhibiting nonlinear dissipation. These insights pave the way for the design of quantum thermal devices and the controlled manipulation of energy transport at the microscopic scale.


    \vspace{1em}
    \noindent\textbf{Data availability statement} \\
     The numerical simulation codes and data used in this study are available at ~\cite{CodeRepo}.


 \widetext

	\appendix

{\color{black}
\section{Transient heat currents and rectification}
\label{app:transients}

In this Appendix, we analyze the transient heat currents for the purely linear and nonlinear dissipation scenarios considered in Sec. \ref{sec:results} (one–photon, and two–photon). We begin with the linear (one–photon) reference case.\\ \\
{\it{Linear case}}:
\label{app:transient_linear}
Before turning to nonlinear dissipation, it is useful to fix the linear case as a reference not only at steady state (Eq.~\eqref{eq:JR_1}) but also during the approach to stationarity.
In the purely linear setup, the mean occupation $\langle \hat{n}\rangle \equiv n$ satisfies a single–rate equation,
\begin{equation}
\dot n (t)
=-\gamma\big(  n (t)-n_0\big),
\label{eq:app_dn_linear}
\end{equation}
where \(n_0\) is the steady-state value of \( n(t)\) for the linear case as given in Eq. \eqref{eq:n0}, and $\gamma = \gamma_L +\gamma_R$.
Its solution is
\begin{equation}
n(t)=
n_0+\big( n(0)-n_0\big)\,e^{-\gamma t},
\label{eq:app_n_sol_linear}
\end{equation}
so the right–bath current relaxes monotonically and exponentially to Eq.~\eqref{eq:JR_1}. Using Eqs. \eqref{eq:app_n_sol_linear} and \eqref{eq:HeatCurrentDef}, we obtain
\begin{align}
\mathcal{J}_R(t)
&=\mathcal{J}_R^{\rm ss}
+\omega\,\gamma_R\big(n_0- n(0)\big)\,e^{-\gamma t},
\label{eq:JR_linear}
\end{align}
with a single time constant $\tau=1/\gamma$. If the initial current has the opposite sign to the steady-state value \(\mathcal{J}_R^{\rm ss}\), the transient crosses zero once at
\begin{equation}
t_*=\frac{1}{\gamma}\,
\ln\!\left(\frac{n_0-  n(0)}{\,n_0-\bar n_R\,}\right),
\label{eq:app_tstar}
\end{equation}
otherwise, the sign never flips. Equation~\eqref{eq:app_tstar} follows by setting \(\mathcal{J}_R(t_*)=0\), i.e. \( n(t_*)=\bar n_R\), and solving Eq.~\eqref{eq:app_n_sol_linear} for \(t_*\). This provides a simple benchmark for the nonlinear cases analyzed next. \\ \\
{\it Nonlinear case}:
We analyze the purely nonlinear setting where only two–photon exchange with the baths is present, i.e., all one–photon rates are set to zero (\(\gamma_\alpha=0\)) while two–photon rates remain finite (\(\Gamma_\alpha>0\)).
Because parity is conserved and the dynamics decomposes into even and odd ladders; we focus on the even sector relevant for two–photon exchange.

To obtain closed formulas for the transient heat current, we consider the low–temperature regime \(T_\alpha \ll 2\omega\), so that the two–photon thermal occupancy
\(\bar m_\alpha\) satisfies \(\bar m_\alpha\ll1\) and populations above \(n=2\) are exponentially suppressed.
Under these conditions we truncate the even ladder to \(\{|0\rangle,|2\rangle\}\) with probabilities \(P_0(t)\) and \(P_2(t)\), \(P_0(t)+P_2(t)\simeq1\).
Projecting the two–photon master equation \eqref{eq:fullmaster} onto \(|2\rangle\!\langle2|\) yields
\begin{equation}
\label{eq:P2_final}
\dot P_2(t)=-\kappa\,P_2(t)+k_{\uparrow},\qquad P_0(t)=1-P_2(t),
\end{equation}
with total upward and downward rates
\begin{equation}
\label{eq:rates_totals}
k_{\uparrow}= \sum_\alpha k_{\uparrow,\alpha}=2\sum_\alpha \Gamma_\alpha \bar m_\alpha,\qquad
k_{\downarrow}= \sum_\alpha k_{\downarrow,\alpha}=2\sum_\alpha \Gamma_\alpha (\bar m_\alpha+1),
\end{equation}
and total relaxation rate \(\kappa=k_{\downarrow}+k_{\uparrow}\).
The factor of \(2\) in \eqref{eq:rates_totals} follows from the matrix elements
\(|\langle0|a^{2}|2\rangle|^2=|\langle2|a^{\dagger2}|0\rangle|^2=2\). Solving Eq.~\eqref{eq:P2_final} yields a single–exponential relaxation,
\begin{equation}
\label{eq:P2_solution}
P_2(t)=P_2^{\mathrm{ss}}+(P_2(0)-P_2^{\mathrm{ss}})e^{-\kappa t},
\qquad
P_2^{\mathrm{ss}}=\frac{k_{\uparrow}}{\kappa},
\end{equation}
with a similar time constant \(1/\kappa\) as the linear (one–photon) benchmark.
The right–bath heat current for two–photon exchange with two-level approximation is given by
\begin{equation}
\label{eq:JR_def_2ph}
\mathcal{J}_R(t)=2\omega(\kappa_R\,P_2(t)-k_{\uparrow,R}),
\end{equation}
where, \(k_{\uparrow,R}=2\Gamma_R\bar m_R\), \(k_{\downarrow,R}=2\Gamma_R(\bar m_R+1)\), and \(\kappa_R =k_{\downarrow,R}+k_{\uparrow,R}\).
Inserting Eq. \eqref{eq:P2_solution} into Eq. \eqref{eq:JR_def_2ph} gives the single–exponential transient
\begin{equation}
\label{eq:JR_t_2ph}
\mathcal{J}_R(t)=\mathcal{J}_R^{\mathrm{ss}}
+2\omega\,\kappa_R(P_2(0)-P_2^{\mathrm{ss}})e^{-\kappa t},
\end{equation}
with the steady state heat current given by
\begin{equation}
\label{eq:JR_ss_2ph}
\mathcal{J}_R^{\mathrm{ss}}
=2\omega\left(\frac{\kappa_R\,k_{\uparrow}}{\kappa}-k_{\uparrow,R}\right).
\end{equation}
Like the linear case, \(\mathcal{J}_R(t)\) given in Eq. \eqref{eq:JR_t_2ph} is strictly monotone and crosses zero at most once; when a sign flip occurs, the zero–crossing time is
\begin{equation}
\label{eq:tstar_2ph}
t_*=\frac{1}{\kappa}\,
\ln\!\left(\frac{\kappa_R P_2(0)-k_{\uparrow,R}}{k_{\uparrow,R}-\kappa_R P_2^{\mathrm{ss}}}\right).
\end{equation}

 \begin{figure*}[t]
  \centering
  \includegraphics[width=\textwidth]{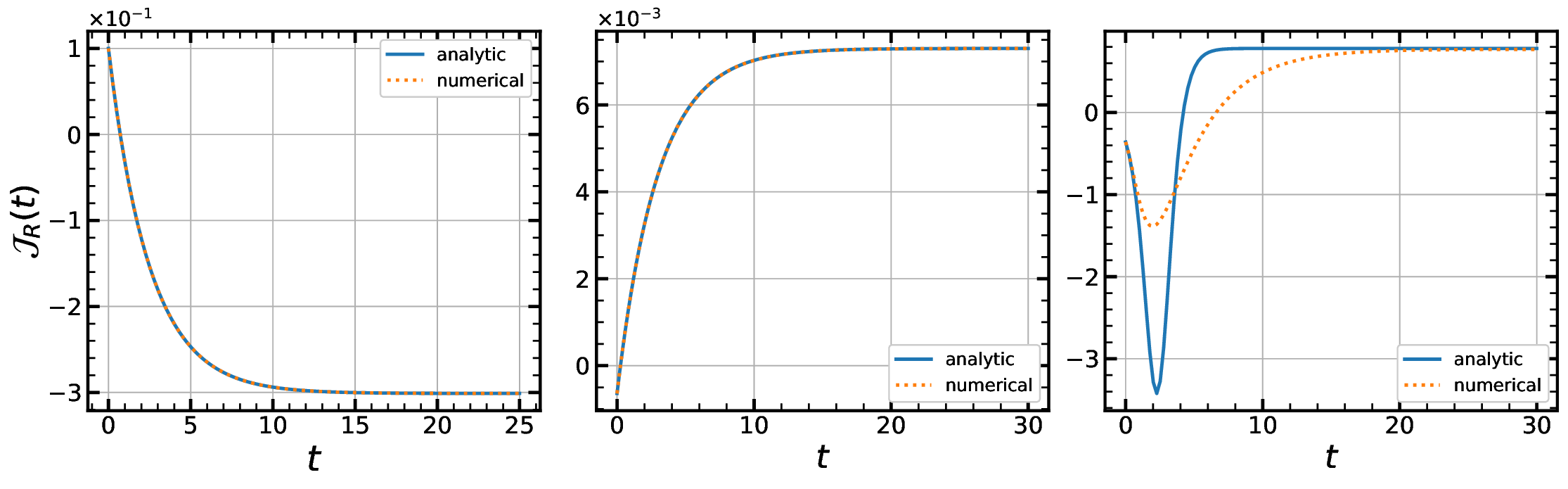}
  \caption{{\color{black}Transient heat current into the right bath, $\mathcal{J}_R(t)$, comparing the analytic results derived in the Appendix \ref{app:transients} with numerical master–equation \eqref{eq:fullmaster} solutions (dotted).  Panels: (a) linear one–photon benchmark with $\omega=1$, $\gamma_L=0.30$, $\gamma_R=0.10$, $T_L=5.5$, and  $T_R=1.45$);  (b) purely nonlinear two–photon, low–$T$ regime with  $\Gamma_L=\Gamma_R=0.08$ and two–photon thermal occupancies  $m_L=0.05$, $m_R=0.002$, compared against the two–level $(|0\rangle,|2\rangle)$ analytic solution; 
  (c) purely nonlinear two–photon, high–$T$ regime with  $\Gamma_L=\Gamma_R=0.01$ and $m_L=10$, $m_R=9$, compared against the high–$T$ closure analytics.  In all cases the initial state is the vacuum, and  the oscillator Hilbert space is truncated to $N=150$.}}
  \label{fig:JR_transients}
\end{figure*}

Complementary to the low-temperature two-level truncation in Eqs.~\eqref{eq:P2_final}–\eqref{eq:tstar_2ph}, we now analyze the opposite regime in which two-photon exchange is strongly thermally activated, \( \bar m_\alpha\gg1 \). It is convenient to work with number moments \( n(t)=\langle \hat n\rangle \) and \( F_2(t)=\langle  n( n-1)\rangle \). From Eq.~\eqref{eq:fullmaster} one obtains the exact balance for the first moment,
\begin{equation}
\dot n(t)=-2\Sigma_\Gamma\,F_2(t)+8A\,n(t)+4A,
\label{eq:n(t)_highT}
\end{equation}
where \( \Sigma_\Gamma\) and \( A \) are given in Eq. \eqref{R4}. The corresponding exact equation for the second factorial moment reads
\begin{equation}
\dot F_2(t)
=\Sigma_\Gamma\!(-4  n(t)^3 +10  n(t)^2- 6  n(t))
+A (24  n(t)^2 + 8  n(t) +4),
\end{equation}
which contains \(n^3\) and thus exposes the usual closure problem for higher moments.
In the high-\(T\) regime the two-photon birth–death process rapidly acquires a geometric profile in each parity sector (the same sector structure responsible for the exact steady-state solution in Sec.~\ref{subsec:IIIB}). Promoting this to a time-local closure yields, with excellent accuracy for \(\bar m_\alpha\gg1\) and exactly at stationarity,
\begin{equation}
F_2(t)\approx 2\,n(t)^2+n(t).
\label{eq:F_2_approx}
\end{equation}
Substituting Eq. \eqref{eq:F_2_approx} into Eq. \eqref{eq:n(t)_highT} gives a closed Riccati equation,
\begin{equation}
\begin{aligned}
\dot n(t)&=-4\Sigma_\Gamma\big(n(t)-n_{\rm ss}\big)\big(n(t)+\tfrac12\big).
\end{aligned}
\end{equation}
With \(n_{\rm ss} =2A/\Sigma_\Gamma\), \(n_{o} = n(0)\), \(K_0 =(n_o-n_{\rm ss})/(n_o+\tfrac12)\), and \(\lambda = 8A+2\Sigma_\Gamma\), the solution is
\begin{equation}
\begin{aligned}
n(t)&=\dfrac{\,n_{\rm ss}+\tfrac12 K_0 e^{-\lambda t}\,}{1-K_0 e^{-\lambda t}}.
\end{aligned}
\end{equation}
The heat current into the right bath follows from the definition used in Eq. \eqref{eq:HeatCurrentDef}. Using the closure given in Eq. \eqref{eq:F_2_approx}, one obtains an explicit polynomial in \(n(t)\),
\begin{equation}
\mathcal J_R(t)=2\omega\Gamma_R\Big(2\,n(t)^2+\big(1-4\bar m_R\big)n(t)-2\bar m_R\Big),
\end{equation}
and hence the steady-state current by substituting \(n_{\rm ss}\),
\begin{equation}
\mathcal J_R^{\rm ss}=2\omega\Gamma_R\Big(2\,n_{\rm ss}^2+\big(1-4\bar m_R\big)n_{\rm ss}-2\bar m_R\Big).
\end{equation}
For intuition it is useful to express the transient relative to \(\mathcal J_R^{\rm ss}\):
\begin{equation}
\mathcal J_R(t)=\mathcal J_R^{\rm ss}
+2\omega\Gamma_R\Big((4n_{\rm ss}+1-4\bar m_R)\,\Delta(t)+2\,\Delta(t)^2\Big),
\label{eq:JR_highT}
\end{equation}
here, $\Delta = n(t) - n_{\rm ss}$.  To quantify how long it takes for the steady state to overturn an initially imposed flow, we use the zero-crossing time \(t_*\): the first solution of \(\mathcal{J}_R(t_*)=0\). Using Eq. \eqref{eq:JR_highT} the zero-crossing time is given by
\begin{equation}
\label{eq:tstar_highT}
t_*=\frac{1}{\lambda}\,
\ln\!\left(
\frac{\big(n_o-n_{\rm ss}\big)\big(n_o^{(R)}+\tfrac{1}{2}\big)}
{\big(n_o^{(R)}-n_{\rm ss}\big)\big(n_o+\tfrac{1}{2}\big)}
\right),
\end{equation}
with
\begin{equation}
\label{eq:n0R}
n_o^{(R)}=\frac{(4\bar m_R-1)+\sqrt{(4\bar m_R-1)^2+16\,\bar m_R}}{4},
\end{equation}
\noindent whenever $\big(n_o-n_0^{(R)}\big)\big(n_{\rm ss}-n_0^{(R)}\big)<0$. If not, there is no crossing; the late-time approach is single-exponential with rate $\lambda$.

Fig.~\ref{fig:JR_transients} compares the analytic transients \(\mathcal{J}_R(t)\) with numerical master equation \eqref{eq:fullmaster} simulations for three regimes:
(a) one–photon (Eq. \eqref{eq:JR_linear}), (b) purely two–photon at low \(T\) (Eq. \eqref{eq:JR_t_2ph}), and (c) purely two–photon at high \(T\) (Eq. \eqref{eq:JR_highT}). In panels (a)–(b) the agreement is essentially perfect because the closed forms are exact (linear) or
controlled by the two–level truncation (low \(T\)). In panel (c) a visible discrepancy appears at early times: the high-\(T\) formula is obtained by a time-local moment closure Eq. \eqref{eq:F_2_approx} that assumes a parity-geometric number distribution. Starting from the vacuum, the distribution is initially far from geometric; higher
cumulants (skewness and kurtosis) are therefore misrepresented, which biases \(\langle a^{\dagger 2}a^{2}\rangle\) and shifts the predicted zero-crossing time \(t_\ast\).
As the distribution relaxes towards geometric within the even ladder, the closure becomes accurate, and both curves approach the same \(\mathcal{J}_R^{\rm ss}\) with the asymptotic rate \(\lambda\) from
Eq.~\eqref{eq:JR_highT}. In addition, numerically reaching the strictly asymptotic high–temperature regime \((\bar m_\alpha \gg 1)\) is demanding: two–photon pumping drives large occupations \(n_{\mathrm{ss}}\sim 2A/\Sigma_\Gamma=O(\bar m)\), so accurate GKLS simulations require very large Hilbert–space cutoffs and stiff integration, which we avoid in panel (c).
Consequently, panel (c) uses only moderately large \(\bar m_\alpha\) rather than the true asymptotic limit, and part of the discrepancy therefore reflects this non–asymptotic choice in addition to the transient moment–closure error.

}

{\color{black}
    \section{Parity-resolved steady state for two-photon dynamics}\label{app:parity}
 
Here we provide a self-contained derivation showing that, for purely two-photon Lindblad dynamics, (i) the even and odd Fock sectors decouple, (ii) within each sector the stationary populations form a normalized geometric law with a common ratio \(r\), and (iii) if the initial state is a mixture of parities, the sector weights simply enter observables as convex coefficients.
We start from the two-photon master equation \eqref{eq:fullmaster} (assuming single-photon decay rate $\gamma_\alpha = 0$) and denote Fock populations by \(P_n(t)=\langle n|\rho(t)|n\rangle\). Using
\(a^2|n\rangle=\sqrt{n(n-1)}\,|n-2\rangle\) and
\(a^{\dagger 2}|n\rangle=\sqrt{(n+1)(n+2)}\,|n+2\rangle\), the diagonal form becomes
\begin{align}
\dot P_n
&=\sum_\alpha \Gamma_\alpha\Big[
(\bar m_\alpha+1)\big((n+2)(n+1)P_{n+2}-n(n-1)P_n\big)
+\bar m_\alpha\big(n(n-1)P_{n-2}-(n+1)(n+2)P_n\big)
\Big].
\label{eq:A_pop}
\end{align}
In this rate-equation only \(n\!\leftrightarrow\!n\pm2\) transitions appear, so parity is conserved: even \(n\)’s never couple to odd \(n\)’s and vice-versa.
At stationarity (\(\dot P_n=0\)), we define the total downward and upward two-photon rates
\begin{equation}
\mathcal A=\sum_\alpha \Gamma_\alpha(\bar m_\alpha+1),\qquad
\mathcal B=\sum_\alpha \Gamma_\alpha \bar m_\alpha .
\label{eq:A_AB}
\end{equation}
For each link \(n\leftrightarrow n+2\), the probability current must vanish. From \eqref{eq:A_pop} this yields
\begin{equation}
\mathcal A\,(n+2)(n+1)P_{n+2}-\mathcal B\,(n+1)(n+2)P_n=0,
\label{eq:A_link}
\end{equation}
and since \((n+1)(n+2)>0\) we obtain the same constant ratio for every link,
\begin{equation}
\frac{P_{n+2}}{P_n}=r,
\qquad
r=\frac{\mathcal B}{\mathcal A}\in(0,1).
\label{eq:A_r}
\end{equation}
We split the chain into parity subsequences
\begin{equation}
q_k=P_{2k}\quad(\text{even}),\qquad s_k=P_{2k+1}\quad(\text{odd}),\qquad k=0,1,2,\ldots
\label{eq:A_parity_defs}
\end{equation}
and apply Eq. \eqref{eq:A_r} with \(n=2k\) and \(n=2k+1\). Both satisfy identical first-order recurrences,
\begin{equation}
q_{k+1}=r\,q_k,\qquad s_{k+1}=r\,s_k,
\label{eq:A_rec}
\end{equation}
whose general solutions are geometric,
\begin{equation}
q_k=C_{\rm e}\,r^k,\qquad s_k=C_{\rm o}\,r^k,
\qquad (C_{\rm e},C_{\rm o}\ge0).
\label{eq:A_geo}
\end{equation}
Let the total (conserved) weights in the even and odd sectors be
\begin{equation}
w_{\rm e}=\sum_{k\ge0}q_k=\sum_{k\ge0}P_{2k},\qquad
w_{\rm o}=\sum_{k\ge0}s_k=\sum_{k\ge0}P_{2k+1},
\qquad w_{\rm e}+w_{\rm o}=1.
\label{eq:A_weights}
\end{equation}
Using \(\sum_{k\ge0}r^k=1/(1-r)\), normalization within each sector fixes the constants in \eqref{eq:A_geo}:
\begin{equation}
C_{\rm e}=w_{\rm e}(1-r),\qquad C_{\rm o}=w_{\rm o}(1-r).
\label{eq:A_Cs}
\end{equation}
Therefore, the unconditional stationary populations are
\begin{equation}
P_{2k}=w_{\rm e}(1-r)\,r^k,\qquad
P_{2k+1}=w_{\rm o}(1-r)\,r^k,
\label{eq:A_uncond}
\end{equation}
and, conditioning on parity (divide by \(w_{\rm e}\) or \(w_{\rm o}\)),
\begin{equation}
\tilde q_k=\frac{P_{2k}}{w_{\rm e}}=(1-r)\,r^k,\qquad
\tilde s_k=\frac{P_{2k+1}}{w_{\rm o}}=(1-r)\,r^k,
\qquad
\sum_{k\ge0} \tilde q_k=\sum_{k\ge0} \tilde s_k=1.
\label{eq:A_cond}
\end{equation}
Hence, both even and odd sectors admit the same normalized geometric law with common ratio \(r\); the only difference in the full state is the overall sector weights \(w_{\rm e},w_{\rm o}\).
Finally, for any function \(f(n)\), the steady-state average decomposes as a convex combination of parity-conditioned averages,
\begin{equation}
\langle f(n)\rangle
=\sum_{n\ge0} f(n)P_n
=w_{\rm e}\sum_{k\ge0} f(2k)\,\tilde q_k
+w_{\rm o}\sum_{k\ge0} f(2k+1)\,\tilde s_k.
\label{eq:A_avg}
\end{equation}
A useful identity that follows immediately from \eqref{eq:A_r} is
\begin{equation}
\langle n(n-1)\rangle
= r\,\langle (n+1)(n+2)\rangle,
\label{eq:A_identity}
\end{equation}
which holds in either sector and for any parity mixture. This relation, together with \(r\) from Eq. \eqref{eq:A_r}, is precisely used to derive the steady-state analytical equation \eqref{eq:AnaHeatTwo} for the right bath heat current.

\medskip
{\it{Steady-state right-bath current by parity}}:
The right-bath heat current for the two-photon model is given by Eq. \eqref{eq:RHCurrent_2ss}. Using the parity-conditioned geometric laws, the factorial moments for odd sector is given by
\begin{align}
\langle n(n{-}1)\rangle_{\rm o}=\frac{2r(3+r)}{(1-r)^2},\qquad\ \
\langle (n{+}1)(n{+}2)\rangle_{\rm o}=\frac{2(3+r)}{(1-r)^2}.
\label{eq:app_moments_odd}
\end{align}
They satisfy the sector-independent identity \eqref{eq:A_identity}.
Substituting Eq.~\eqref{eq:app_moments_odd} into Eq.~\eqref{eq:RHCurrent_2ss}, we obtain
\begin{align}
\mathcal J_R^{\rm odd}
&= \frac{4\omega\,\Gamma_R}{\Sigma_\Gamma^2}\,
\big(\Sigma_\Gamma \bar m_R- A\big)\,\big(3\Sigma_\Gamma+4 A\big).
\label{eq:app_JR_odd}
\end{align}
If the initial state has parity weights \(w_{\rm e},w_{\rm o}\),
parity is conserved and the steady state is the convex mixture of the two ladders. Since
Eq. \eqref{eq:RHCurrent_2ss} is linear in \(\hat \rho_{\rm ss}\), the current is the convex
combination
\begin{equation}
\mathcal J_R^{\rm mix}
= w_{\rm e}\,\mathcal J_R^{\rm even} + w_{\rm o}\,\mathcal J_R^{\rm odd}
= \frac{4\omega\,\Gamma_R}{\Sigma_\Gamma^2}\,\big(\Sigma_\Gamma \bar m_R- A\big)
\Big(\,4 A + \big(1+2w_{\rm o}\big)\Sigma_\Gamma\Big).
\label{eq:app_JR_mixed}
\end{equation}
In Sec. \ref{subsec:IIIB} we initialize in the even sector, \(w_{\rm e}=1\), so the two-photon-only
closed-form current refers to Eq.~\eqref{eq:AnaHeatTwo}.
}

\begin{figure*}[!t]
  \centering
  \includegraphics[width=0.98\textwidth]{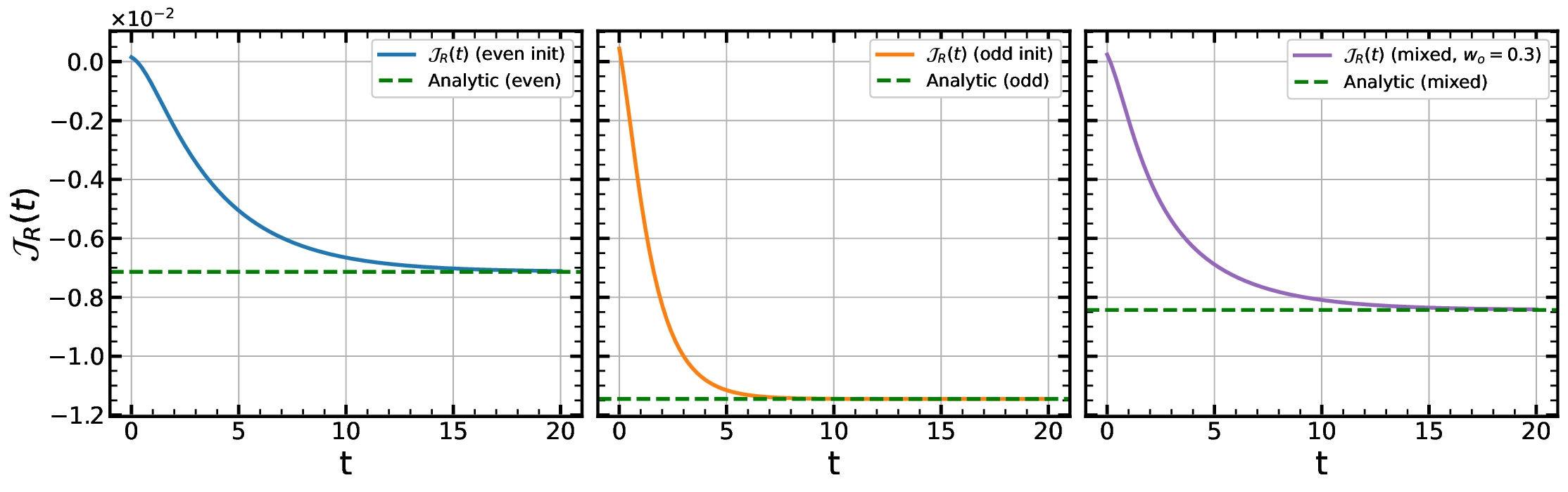}
  \caption{{\color{black}Time-dependent right-bath heat current \(\mathcal{J}_R(t)\) in the two-photon–only model, obtained by numerically integrating the master equation~\eqref{eq:fullmaster} with single-photon dissipation set to zero \((\gamma_\alpha = 0)\). Left:  even-sector initialization \(|0\rangle\). Middle: odd-sector initialization \(|1\rangle\). Right: mixed parity with odd weight \(w_{\rm o}=0.3\). Dashed lines are the corresponding steady-state analytical currents:
  even sector \(\mathcal{J}_R^{\rm even}\) (Eq.~\eqref{eq:AnaHeatTwo}), odd sector \(\mathcal{J}_R^{\rm odd}\) (Eq.~\eqref{eq:app_JR_odd}),  and mixed parity \(\mathcal{J}_R^{\rm mix}\) (Eq.~\eqref{eq:app_JR_mixed}), confirming convergence to the parity-resolved predictions derived in  Appendix~\ref{app:parity}. Parameters: \(\omega=1\), \(\Gamma_L=0.1\),
  \(\Gamma_R=10^{-3}\), \(T_L=2\) and \(T_R=0.6\),  and cutoff \(n_{\max}=40\).}}
  \label{fig:parity_currents}
\end{figure*}

	\section{Enhanced thermal rectification via higher-order photon processes}\label{sec:Appendix A}

    In this appendix, we investigate thermal rectification induced by photon exchange processes with the thermal baths that go beyond two-photon interactions, focusing specifically on the case of three-photon dissipation. We show that the degree of thermal rectification systematically increases with the order of the photon exchange process. This enhancement is particularly significant at low temperatures, where the thermal blockade mechanism becomes increasingly pronounced. Higher-order processes require the harmonic oscillator to populate progressively higher Fock states to engage in energy exchange. In the presence of a cold bath, the oscillator is driven into low-occupancy states, thereby suppressing multi-photon transitions and creating a stronger directional bias in heat flow. These findings extend the rectification mechanism identified for two-photon processes and highlight the growing effectiveness of nonlinear dissipation in enabling directional heat transport as the process order increases.

    To model the effect of higher-order dissipation, we extend the system-bath interaction to include three-photon exchange processes in addition to the previously considered single- and two-photon channels. Within the Born-Markov and secular approximations, the reduced dynamics of the harmonic oscillator are governed by the Lindblad master equation:
    \begin{equation}\label{eq:full3photon}
    \frac{d\hat{\rho}}{dt} = -i[\hat{H}_S, \hat{\rho}] + \sum_{\alpha = L, R} \left( \mathcal{L}^{(1)}_{\alpha}[\hat{\rho}] + \mathcal{L}^{(2)}_{\alpha}[\hat{\rho}] + \mathcal{L}^{(3)}_{\alpha}[\hat{\rho}] \right),
    \end{equation}
    where $\hat{H}_S = \omega \hat{a}^\dagger \hat{a}$ is the system Hamiltonian, and the dissipators $\mathcal{L}^{(k)}_\alpha[\hat{\rho}]$ describe $k$-photon absorption and emission processes mediated by bath $\alpha$. The three-photon dissipator takes the form:
    \begin{equation}
    \mathcal{L}^{(3)}_{\alpha}[\hat{\rho}] = \gamma^{(3)}_\alpha (\bar{n}^{(3)}_\alpha + 1)\, \mathcal{D}[\hat{a}^3]\,\hat{\rho} + \gamma^{(3)}_\alpha \bar{n}^{(3)}_\alpha \, \mathcal{D}[\hat{a}^{\dagger 3}]\,\hat{\rho},
    \end{equation}
    with $\bar{n}^{(3)}_\alpha = (e^{3\omega/T_\alpha} - 1)^{-1}$ denoting the thermal occupation at frequency $3\omega$, and $\gamma^{(3)}_\alpha$ the three-photon dissipation rate. The structure of the dissipator reflects the fact that energy exchange through this channel requires transitions between Fock states separated by three quanta.

    To evaluate the energy flux from the right reservoir, we compute the steady-state heat current as the average rate of energy change in the system due to dissipative processes induced by the right bath. This includes contributions from single-, two-, and three-photon exchange mechanisms. The total steady-state heat current from the right bath is given by:
    \begin{equation}
    \mathcal{J}_R = \mathrm{Tr} \left[ \hat{H}_S \left( \mathcal{L}^{(1)}_R + \mathcal{L}^{(2)}_R + \mathcal{L}^{(3)}_R \right)(\hat{\rho}_{\mathrm{ss}}) \right].
        \end{equation}

\begin{figure}[!htbp]
		\begin{center}
			\leavevmode
			\includegraphics[width=0.48\textwidth,angle=0]{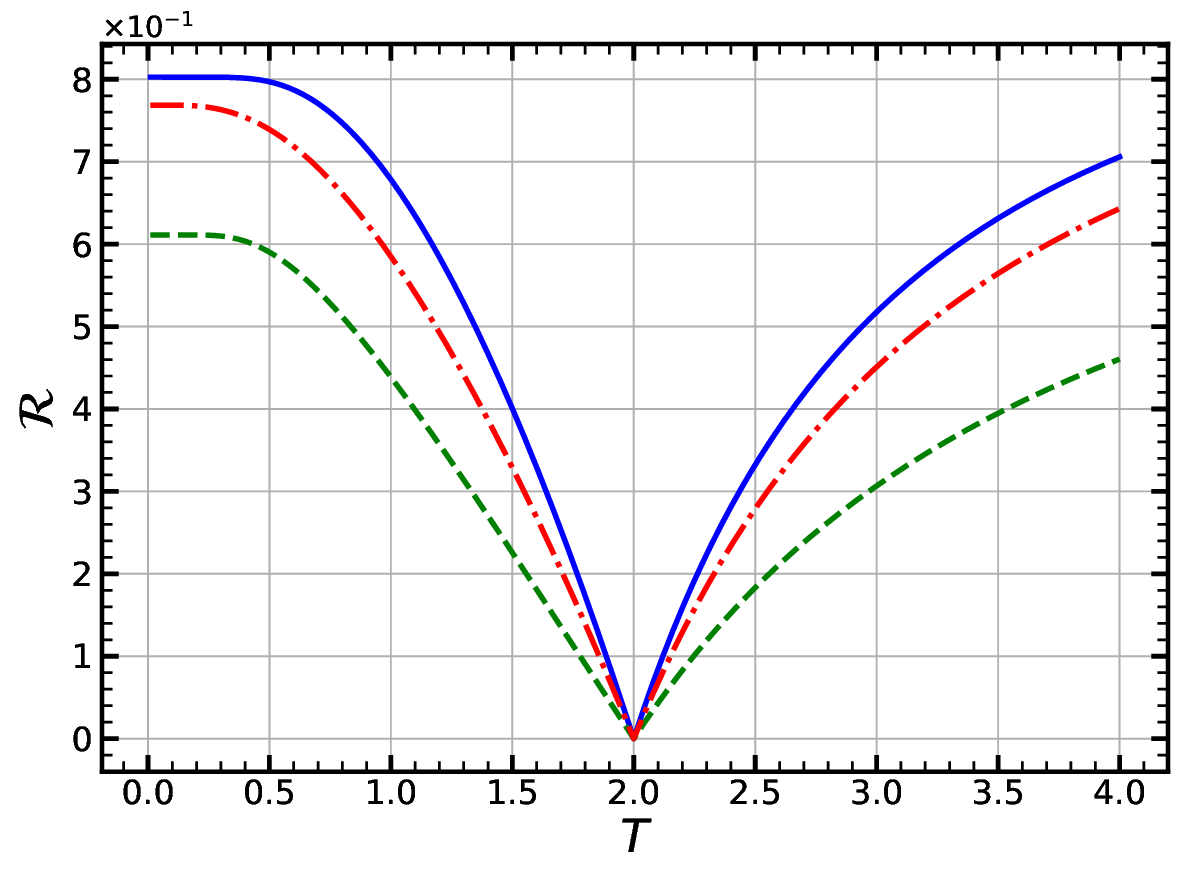}
			\caption{Rectification coefficient \( \mathcal{R} \) as a function of bath temperature \( T \) for different multiphoton dissipation scenarios. The solid blue curve corresponds to three-photon exchange only (\( \gamma^{(3)}_L = 0.1 \), \( \gamma^{(3)}_R = 0.01 \)); the dashed green curve corresponds to two-photon exchange only (\( \gamma^{(2)}_L = 0.1 \), \( \gamma^{(2)}_R = 0.01 \)); and the red dash-dotted curve includes all three dissipation channels with \( \gamma^{(1)}_L = 0.1 \), \( \gamma^{(1)}_R = 0.01 \), \( \gamma^{(2)}_L = 0.1 \), \( \gamma^{(2)}_R = 0.01 \), \( \gamma^{(3)}_L = 0.1 \), and \( \gamma^{(3)}_R = 0.01 \). The left bath temperature is fixed at \( T_L = 2.0 \), while the right bath temperature \( T_R \) is varied. The rectification coefficient is computed by numerically integrating the master equation~\eqref{eq:full3photon} and evaluating the heat current flowing into the right bath in both forward and reverse configurations.}\label{fig:fig6}
		\end{center}
	\end{figure}

    In the presence of multiphoton processes, obtaining an analytical solution becomes intractable. Therefore, we compute the steady-state heat current by numerically integrating the master equation~(\ref{eq:full3photon}) and evaluate the corresponding rectification. The numerical results are presented in Fig.~\ref{fig:fig6}, which shows the rectification coefficient \( \mathcal{R} \) as a function of bath temperature \( T \) for various multiphoton dissipation configurations. The most prominent feature is that rectification is significantly enhanced in the case of three-photon processes compared to both the two-photon-only case and the scenario where all dissipation channels are present. This demonstrates that the degree of rectification increases systematically with the order of the photon exchange process, highlighting the growing asymmetry induced by nonlinear interactions.

    The enhanced rectification observed in the three-photon case at low temperatures can be attributed to a stronger thermal blockade mechanism, as discussed in Subsec.~\ref{subsec:IIIB}. When the cold bath is strongly coupled, it drives the harmonic oscillator into low-occupancy Fock states. Since higher-order photon emission processes (e.g., three-photon) require the occupation of higher-energy states, their contribution to heat transport is effectively suppressed under these conditions. This blockade becomes increasingly pronounced for higher-order dissipation, leading to a stronger suppression of heat flow in one direction. When the thermal bias is reversed, such that the cold bath is weakly coupled, this blockade is lifted, allowing the hot, strongly coupled bath to activate the nonlinear transitions. This directional asymmetry results in enhanced thermal rectification.

    At high temperatures, where thermal excitation of higher Fock states becomes more probable, the role of multiphoton processes becomes more prominent. In particular, the efficiency of three-photon energy exchange increases significantly when the hot bath is strongly coupled, thereby enhancing the forward heat current relative to the reverse case. As a result, the rectification remains appreciable even at elevated temperatures, and is greater for three-photon processes than for two-photon ones.

   In the configuration involving all dissipation channels—including single-photon processes—the rectification is noticeably reduced compared to the purely nonlinear cases. This is due to the symmetric nature of single-photon interactions, which allows energy transport in both directions even when the system is in a low-excitation state. As such, single-photon processes provide a linear transport pathway that partially lifts the thermal blockade, diminishing the asymmetry and thus reducing the rectification effect. This underscores the importance of selectively engineering nonlinear dissipation channels to optimize rectification in quantum thermal devices.

    In principle, the three-photon exchange process with the thermal bath can also be implemented using the auxiliary-system scheme described in Sec.~\ref{sec:ExperimentalFeas}, where a TLS mediates the interaction between the harmonic oscillator and the reservoir. To realize this, one must extend the expansion of the operator \( {\sigma}_+(t) \) in Eq.~(\ref{A12}) to third order in the coupling parameter \( \alpha \), retaining terms up to \( \mathcal{O}(\alpha^3) \) and neglecting higher-order contributions. This yields additional terms in the interaction Hamiltonian that involve cubic powers of the ladder operators, leading to three-photon dissipative channels in the effective master equation. As a result, the engineered dissipation not only reproduces one- and two-photon processes but also enables controlled implementation of three-photon interactions. This provides a viable route for experimentally exploring higher-order nonlinear thermal transport effects.

{\color{black}
\section{Error bound for the mean-field closure in the hybrid asymmetric case}
\label{app:errorbound}

We quantify the accuracy of the mean-field closure used in Sec.~\ref{subsec:IIIc} for the hybrid configuration
\((\Gamma_L\neq 0,\ \Gamma_R=0)\). In this setting, the steady-state equation for the mean occupation follows from
Eq.~\eqref{eq:dn_dt}. The closure in Eq.~\eqref{eq:factor} replaces the second-order moments by functions of
\(n\). Here we benchmark that replacement against the exact two-photon steady state from
Sec.~\ref{subsec:IIIB}, using the moments derived in Eqs.~\eqref{eq:average}–\eqref{eq:averages} as a reference.
In the hybrid case with \(\Gamma_R=0\), Eq.~\eqref{eq:rdef} reduces to
\begin{equation}
r= e^{-2\omega/T_L}.
\label{eq:app_r_hybrid}
\end{equation}
Evaluating Eq.~\eqref{eq:factor} at the exact mean given in Eq. \eqref{eq:average} yields
\begin{equation}
\langle  n( n-1)\rangle_{\mathrm{fact}}=\langle  n\rangle^2-\langle{n}\rangle=\frac{2r(3r-1)}{(1-r)^2}, \qquad\qquad
\langle ( n+1)( n+2)\rangle_{\mathrm{fact}}=\langle {n} \rangle^2 + 3\langle {n} \rangle + 2=\frac{2(1+r)}{(1-r)^2}.
\label{eq:n_fact}
\end{equation}
To measure the closure error on the second-order moments, we introduce the relative deviations
\begin{equation}
\varepsilon_2^{(-)}(r)=
\frac{\langle  n( n-1)\rangle_{\mathrm{fact}}-\langle  n( n-1)\rangle_{\mathrm{exact}}}
{\langle  n( n-1)\rangle_{\mathrm{exact}}},
\qquad
\varepsilon_2^{(+)}(r)=
\frac{\langle ( n+1)( n+2)\rangle_{\mathrm{fact}}-\langle ( n+1)( n+2)\rangle_{\mathrm{exact}}}
{\langle ( n+1)( n+2)\rangle_{\mathrm{exact}}}.
\end{equation}
Using Eqs.~\eqref{eq:average}–\eqref{eq:averages} together with \eqref{eq:n_fact}, straightforward algebra gives
\begin{equation}
\ \varepsilon_2^{(-)}(r)=-\frac{2}{1+3r},\qquad
\varepsilon_2^{(+)}(r)=-\frac{2r}{1+3r}\ .
\label{eq:app_epsilons}
\end{equation}
Thus, in this regime, the factorization underestimates both factorial moments.
The two-photon bracket appearing in Eq.~\eqref{eq:dn_dt} is
\begin{equation}
E =\bar m_L\langle( n+1)( n+2)\rangle-(\bar m_L+1)\langle  n( n-1)\rangle.
\label{eq:E}
\end{equation}
We denote the closure-induced discrepancy by \(\Delta E=E_{\mathrm{fact}}-E_{\mathrm{exact}}\), i.e., the difference between the value of \(E\) computed with the factorization in Eq.~\eqref{eq:factor} and its exact evaluation. Noting \(E_{\mathrm{exact}}=0\), and using Eq.~\eqref{eq:n_fact} together with Eq.~\eqref{eq:E}, one obtains
\begin{equation}
\ \Delta E=\frac{4r}{(1-r)^2}\, ,
\label{eq:app_deltaE}
\end{equation}
which depends on \(T_L\) only through \(r\) in Eq.~\eqref{eq:app_r_hybrid}.
We now recast Eq.~\eqref{eq:dn_dt} at steady state using Eq.~\eqref{eq:E}:
\begin{equation}
0=\gamma\,(n_0-n)+2\Gamma_L\,E.
\label{eq:dn/dt=0}
\end{equation}
Let \(n^\star\) be the solution of Eq.~\eqref{eq:dn/dt=0} with \(E=E_{\mathrm{exact}}\), and \(n_{\mathrm{fact}}\) the solution with
\(E=E_{\mathrm{fact}}\). Linearizing around \(n^\star\) gives
\begin{equation}
-\gamma\,(n_{\mathrm{fact}}-n^\star)+2\Gamma_L\,\Delta E=0,
\end{equation}
and therefore the absolute error in the mean is
\begin{equation}
\ n_{\mathrm{fact}}-n^\star=\frac{8\Gamma_L}{\gamma}\,\frac{r}{(1-r)^2}\ .
\label{eq:app_deltan}
\end{equation}
Using Eq.~\eqref{eq:heatsubC}, Eq.~\eqref{eq:app_deltan} implies the following bound on the current error:
\begin{equation}
\ |\delta \mathcal J_R| \;=\; \ | \mathcal J_R^{\mathrm{fact}} \;-\; \mathcal J_R^{\mathrm{exact}} | \le
\omega\,\gamma_R\,\frac{8\Gamma_L}{\gamma}\,\frac{r}{(1-r)^2}\,.
\label{eq:app_curr_bound}
\end{equation}
For a conservative and fully analytic criterion, we compare \(|\delta \mathcal J_R|\) with the linear-bath
current scale from Eq.~\eqref{eq:JR_1}. Imposing \(|\delta \mathcal J_R|\le 0.1\,\mathcal J_{\mathrm{lin}}\) and using
Eq.~\eqref{eq:app_curr_bound} yields
\begin{equation}
\ \Gamma_L\ \le\  \frac{\gamma}{80}\,|\bar n_R-\bar n_L|\,
\frac{\big(1-e^{-2\omega/T_L}\big)^2}{e^{-2\omega/T_L}}\ .
\label{eq:app_10percent}
\end{equation}
Consequently, for fixed \(\gamma_{L,R}\) and thermal bias
\(|\bar n_R-\bar n_L|\), the mean-field closure remains within \(\sim 10\%\) whenever
\(\Gamma_L\) satisfies Eq.~\eqref{eq:app_10percent}.}

{\color{black}
\section{Microscopic derivation of the master equation}\label{App:B}

We consider an ancillary TLS of transition frequency $\omega_0$ with Pauli operators $\sigma_\pm,\sigma_z$, and a single harmonic mode of frequency $\omega_a$ with bosonic operators $a,a^\dagger$. The longitudinal coupling strength is $g$, and we introduce the small, dimensionless dressing parameter $\alpha=g/\omega_a$ such that $\alpha\ll 1$.  Throughout Appendix~\ref{App:B}, for notational convenience, we omit hats over operators (e.g., $\hat a \to a$, $\hat\sigma_x \to \sigma_x$); all such symbols still denote operators. We consider a single bosonic bath with Hamiltonian
\begin{equation}
H_B=\sum_{k}\,\omega_k\,c_k^\dagger c_k,
\label{A1}
\end{equation}
and we couple the TLS transversely to the bath quadrature through
\begin{equation}
H_{SB}=\sigma_x\,(b+b^\dagger),
\qquad
b:=\sum_{k}g_k\,c_k,
\label{A2}
\end{equation}
so that $b$ is the usual collective bath annihilation operator built from the normal modes $c_k$. The full Hamiltonian is therefore $H=H_S+H_B+H_{SB}$, where the system Hamiltonian reads
\begin{equation}
H_S=\frac{\omega_0}{2}\,\sigma_z+\omega_a\,a^\dagger a+g\,\sigma_z\,(a+a^\dagger).
\label{A3}
\end{equation}
Our starting point is the Born–Markov second–order perturbative von–Neumann equation for the reduced density operator for the joint HO-TLS in the interaction picture with respect to $H_S+H_B$. Writing $X(t)$ for the system coupling operator and $B(t)$ for the bath coupling operator in this picture, the single–bath evolution is \cite{Breuer2002}
\begin{align}
\frac{d\rho(t)}{dt}
=-\int_{0}^{\infty}\!ds\;\Big(
C(s)\,[X(t),\,X(t-s)\,\rho(t)] 
+C(-s)\,[X(t),\,\rho(t)\,X(t-s)]
\Big),
\label{A4}
\end{align}
where $B(t)=b(t)+b^\dagger(t)$, $X(t)$ is specified below (see Eq. (\ref{A11})), and
\begin{equation}
    C(s)= \mathrm{Tr}_B\!\big(B(t)\,B(t{-}s)\,\rho_B\big) = \big\langle B(s)B(0)\big\rangle_{\rho_B}
=\big\langle \big(b(s)+b^\dagger(s)\big)\big(b(0)+b^\dagger(0)\big)\big\rangle_{\rho_B}
\end{equation}
is the stationary bath correlation function with respect to some fixed bath state $\rho_B$. The $C(s)$ by stationarity (\([H_B,\rho_B]=0\)), depends only on the time difference \(s\), and 
for Hermitian \(B\) one has \(C(-s)=C(s)^{\ast}\). Note that, to derive Eq. (\ref{A4}) we use the Born approximation (factorized state \(\rho_{\mathrm{tot}}(t)\approx \rho(t)\otimes\rho_B\)),
the Markov approximation (replace \(\rho(t{-}s)\to \rho(t)\) and extend the upper integration limit to \(\infty\)),
and assume a zero bath mean \(\langle B\rangle=\mathrm{Tr}_B(B\rho_B)=0\).

Equation~(\ref{A4}) is the master kernel into which we shall substitute the dressed interaction–picture decomposition of $X(t)$.
To obtain $X(t)$, we first diagonalize the longitudinal interaction term in Eq. (\ref{A3}) by a $\sigma_z$–conditioned displacement polaron transformation. Defining
\begin{equation}
U=\exp\!\Big[-\alpha\,(a^\dagger-a)\,\sigma_z\Big],
\qquad
\tilde{a}:=U^\dagger a\,U=a-\alpha\,\sigma_z,
\label{A5}
\end{equation}
a straightforward calculation yields \cite{PhysRevE.90.022102}
\begin{equation}
U^\dagger H_S U=\frac{\omega_0}{2}\,\sigma_z+\omega_a\,\tilde{a}^\dagger\tilde{a}-\frac{g^2}{\omega_a}.
\label{A6}
\end{equation}
Discarding the irrelevant constant, the free Hamiltonian that generates the system interaction picture is therefore
\begin{equation}
\tilde{H}_0=\frac{\omega_0}{2}\,\sigma_z+\omega_a\,\tilde{a}^\dagger\tilde{a}.
\label{A7}
\end{equation}
In the same dressed frame, the TLS ladder operators acquire oscillator displacements according to
\begin{equation}
\tilde{\sigma}_\pm:=U^\dagger\sigma_\pm U
=\sigma_\pm\,e^{\pm \alpha(\tilde{a}^\dagger-\tilde{a})},
\label{A8}
\end{equation}
Note that $\tilde a^\dagger - \tilde a = a^\dagger - a$, so the exponent in Eq.~(\ref{A8}) can be written with either $a$ or $\tilde a$. From now on we work in the interaction picture generated by $\tilde{H}_0$ for the system and by $H_B$ for the bath. The dressed mode thus rotates as
\begin{equation}
\tilde{a}(t)=\tilde{a}\,e^{-i\omega_a t},
\qquad
\tilde{a}^\dagger(t)=\tilde{a}^\dagger e^{+i\omega_a t},
\label{A9}
\end{equation}
and the dressed TLS operators rotate as
\begin{equation}
\tilde{\sigma}_\pm(t)=\tilde{\sigma}_\pm\,e^{\pm i\omega_0 t}.
\label{A10}
\end{equation}
Because $\tilde{\sigma}_\pm$ contains the displacement exponentials in Eq.~(\ref{A8}), their time dependence includes harmonics of the oscillator frequency. To make this explicit and to prepare a frequency–resolved decomposition for substitution into Eq.~(\ref{A4}), we expand the exponentials in Eq.~(\ref{A8}) to second order in the small parameter $\alpha$. Introducing
\begin{equation}
X_1(t) =\tilde{a}^\dagger e^{+i\omega_a t}-\tilde{a}\,e^{-i\omega_a t},
\label{A11}
\end{equation}
we obtain, up to $\mathcal{O}(\alpha^2)$,
\begin{align}
\tilde{\sigma}_+(t)&=\sigma_+ e^{+i\omega_0 t}\Big(1+\alpha\,X_1(t)+\tfrac{\alpha^2}{2}\,X_1(t)^2\Big)+\mathcal{O}(\alpha^3),
\qquad
\tilde{\sigma}_-(t)=\tilde{\sigma}_+(t)^\dagger,
\label{A12}
\end{align}
with the identity
\begin{equation}
X_1(t)^2=\tilde{a}^{\dagger 2}e^{+2i\omega_a t}+\tilde{a}^{2}e^{-2i\omega_a t}-\big(2\,\tilde{a}^\dagger\tilde{a}+1\big).
\label{A13}
\end{equation}
Equations~(\ref{A9})–(\ref{A13}) show that the dressed TLS operators carry Fourier components at the Bohr frequencies $\omega\in\{\omega_0,\;\omega_0\pm\omega_a,\;\omega_0\pm 2\omega_a\}$. 
We now express the system coupling operator in the dressed interaction picture as
\begin{equation}
X(t)=\tilde{\sigma}_+(t)+\tilde{\sigma}_-(t)
=\sum_{\omega\in\{\omega_0,\ \omega_0\pm\omega_a,\ \omega_0\pm 2\omega_a\}}
\Big(A_\omega\,e^{+i\omega t}+A_\omega^\dagger\,e^{-i\omega t}\Big),
\label{A14}
\end{equation}
where, keeping all terms up to second order in $\alpha$ and using Eqs.~(\ref{A9})--(\ref{A13}), the eigenoperators (jump operators) are
\begin{equation}
\begin{aligned}
A_{\omega_0}&=\sigma_-(1-\tfrac{\alpha^2}{2}-\alpha^2\,\tilde{a}^\dagger\tilde{a}),\nonumber\\
A_{\omega_0+\omega_a}&=\alpha\,\sigma_-\,\tilde{a},\qquad
A_{\omega_0-\omega_a}=-\,\alpha\,\sigma_-\,\tilde{a}^\dagger,\qquad
A_{\omega_0+2\omega_a}=\tfrac{\alpha^2}{2}\,\sigma_-\,\tilde{a}^{2},\qquad
A_{\omega_0-2\omega_a}=\tfrac{\alpha^2}{2}\,\sigma_-\,\tilde{a}^{\dagger 2},
\end{aligned}
\label{A15}
\end{equation}
and $A_{-\omega}=A_\omega^\dagger$.
Substituting the decomposition (\ref{A14}) into the kernel (\ref{A4}) requires the explicit evaluation of the operator products $X(t)X(t-s)$ and $X(t-s)X(t)$ and the two double commutators therein. Using (\ref{A14}) twice, the product $X(t)X(t\!-\!s)$ expands as
\begin{align}
X(t)X(t\!-\!s)
&=\sum_{\omega,\omega'}\Big(
A_\omega A_{\omega'}\,e^{+i\omega t}e^{+i\omega'(t-s)}
+ A_\omega A_{\omega'}^\dagger\,e^{+i\omega t}e^{-i\omega'(t-s)}
+ A_\omega^\dagger A_{\omega'}\,e^{-i\omega t}e^{+i\omega'(t-s)}
+ A_\omega^\dagger A_{\omega'}^\dagger\,e^{-i\omega t}e^{-i\omega'(t-s)}
\Big),
\end{align}
and an analogous expression holds for $X(t\!-\!s)X(t)$ with the order of factors exchanged and $t\leftrightarrow t\!-\!s$ in the phases. Inserting these into the first commutator in (\ref{A4}) gives
\begin{align}
[X(t),\,X(t\!-\!s)\rho(t)]
&=X(t)X(t\!-\!s)\rho(t)-X(t\!-\!s)\rho(t)X(t)\nonumber\\
&=\sum_{\omega,\omega'}\Big(
A_\omega A_{\omega'}\,e^{+i\omega t}e^{+i\omega'(t-s)}\rho(t)
- A_{\omega'}\,\rho(t)\,A_\omega\,e^{+i\omega'(t-s)}e^{+i\omega t}\nonumber\\
&\quad + A_\omega A_{\omega'}^\dagger\,e^{+i\omega t}e^{-i\omega'(t-s)}\rho(t)
- A_{\omega'}^\dagger\,\rho(t)\,A_\omega\,e^{-i\omega'(t-s)}e^{+i\omega t}\nonumber\\
&\quad + A_\omega^\dagger A_{\omega'}\,e^{-i\omega t}e^{+i\omega'(t-s)}\rho(t)
- A_{\omega'}\,\rho(t)\,A_\omega^\dagger\,e^{+i\omega'(t-s)}e^{-i\omega t}\nonumber\\
&\quad + A_\omega^\dagger A_{\omega'}^\dagger\,e^{-i\omega t}e^{-i\omega'(t-s)}\rho(t)
- A_{\omega'}^\dagger\,\rho(t)\,A_\omega^\dagger\,e^{-i\omega'(t-s)}e^{-i\omega t}
\Big),
\end{align}
and the second commutator in (\ref{A4}) is expanded similarly as
\begin{align}
[X(t),\,\rho(t)X(t\!-\!s)]
&=X(t)\rho(t)X(t\!-\!s)-\rho(t)X(t\!-\!s)X(t)\nonumber\\
&=\sum_{\omega,\omega'}\Big(
A_\omega\,\rho(t)\,A_{\omega'}\,e^{+i\omega t}e^{+i\omega'(t-s)}
- \rho(t)\,A_{\omega'}A_\omega\,e^{+i\omega'(t-s)}e^{+i\omega t}\nonumber\\
&\quad + A_\omega\,\rho(t)\,A_{\omega'}^\dagger\,e^{+i\omega t}e^{-i\omega'(t-s)}
- \rho(t)\,A_{\omega'}^\dagger A_\omega\,e^{-i\omega'(t-s)}e^{+i\omega t}\nonumber\\
&\quad + A_\omega^\dagger\,\rho(t)\,A_{\omega'}\,e^{-i\omega t}e^{+i\omega'(t-s)}
- \rho(t)\,A_{\omega'}A_\omega^\dagger\,e^{+i\omega'(t-s)}e^{-i\omega t}\nonumber\\
&\quad + A_\omega^\dagger\,\rho(t)\,A_{\omega'}^\dagger\,e^{-i\omega t}e^{-i\omega'(t-s)}
- \rho(t)\,A_{\omega'}^\dagger A_\omega^\dagger\,e^{-i\omega'(t-s)}e^{-i\omega t}
\Big).
\end{align}
Each of the eight families of terms above is multiplied, respectively, by either $C(s)$ or $C(-s)$ under the $s$–integral in (\ref{A4}). Using stationarity of the bath, the time integrals reduce to one–sided Fourier transforms of the correlation function. We therefore introduce
\begin{equation}
G(+\omega)=\int_{0}^{\infty}\!ds\,e^{+i\omega s}\,C(s),
\qquad
G(-\omega)=\int_{0}^{\infty}\!ds\,e^{-i\omega s}\,C(-s),
\label{A16}
\end{equation}
so that representative factors such as $\int_0^\infty\!ds\, C(s)\,e^{+i\omega'(t-s)}$ and $\int_0^\infty\!ds\, C(-s)\,e^{-i\omega'(t-s)}$ evaluate, respectively, to $e^{+i\omega' t}G(+\omega')$ and $e^{+i\omega' t}G(-\omega')$. 
With the definitions in (\ref{A16}), the $s$–integrals in (\ref{A4}) can be performed term by term. Collecting the contributions coming from $C(s)$ and $C(-s)$, and using the product expansions written above for the two double commutators, one finds the standard Redfield–type expression
\begin{equation}
\frac{d\rho(t)}{dt}
=\sum_{\omega,\omega'} e^{i(\omega'-\omega)t}\Big\{
G(+\omega')\Big(A_{\omega'}\,\rho(t)\,A_\omega^\dagger - A_\omega^\dagger A_{\omega'}\,\rho(t)\Big)
+G(-\omega')\Big(A_\omega\,\rho(t)\,A_{\omega'}^\dagger - \rho(t)\,A_{\omega'}^\dagger A_\omega\Big)
\Big\}.
\label{A17}
\end{equation}
Since the bath quadrature $B(t)$ is Hermitian, its correlation function obeys $C(-s)=C(s)^\ast$, which implies $G(-\omega)=G(+\omega)^\ast$. It is convenient to separate real and imaginary parts by writing
\begin{equation}
G(+\omega)=\tfrac{1}{2}\,\gamma(\omega)+i\,S(\omega),
\qquad
G(-\omega)=\tfrac{1}{2}\,\gamma(\omega)-i\,S(\omega),
\label{A18}
\end{equation}
where $\gamma(\omega)=2\,\mathrm{Re}\,G(+\omega)\ge 0$ gives dissipative rates and $S(\omega)=\mathrm{Im}\,G(+\omega)$ generates coherent (Lamb–shift) contributions.

At this point the generator Eq. (\ref{A17}) contains oscillatory prefactors $e^{i(\omega'-\omega)t}$. In the secular approximation, valid when $|\omega-\omega'|$ is large compared with the relevant relaxation rates, these rapidly rotating terms average to zero on the coarse–grained time scale, and only the stationary components with $\omega=\omega'$ are retained. 
Invoking the secular selection and discarding Lamb–shift, we write the dressed interaction–picture generator in a purely dissipative form. Since $\mathcal{D}[cL]=|c|^{2}\mathcal{D}[L]$, all scalar powers of $\alpha$ are absorbed into the frequency–resolved rates. 
The resulting master equation is given by
\begin{align}
\frac{d\tilde{\rho}(t)}{dt}
&= \gamma(\omega_0)\,
   \mathcal{D}\!\Big[\Big(1-\tfrac{\alpha^2}{2}-\alpha^2\tilde{a}^\dagger\tilde{a}\Big)\tilde{\sigma}_{-}\Big]\tilde{\rho}(t)
 \;+\; \gamma(-\omega_0)\,
   \mathcal{D}\!\Big[\Big(1-\tfrac{\alpha^2}{2}-\alpha^2\tilde{a}^\dagger\tilde{a}\Big)\tilde{\sigma}_{+}\Big]\tilde{\rho}(t)\nonumber\\
&\quad+\; \alpha^{2}\gamma(\omega_0{+}\omega_a)\,
   \mathcal{D}\!\big[\tilde{\sigma}_{-}\tilde{a}\big]\tilde{\rho}(t)
 \;+\; \alpha^{2}\gamma(-\omega_0{-}\omega_a)\,
   \mathcal{D}\!\big[\tilde{\sigma}_{+}\tilde{a}^{\dagger}\big]\tilde{\rho}(t)\nonumber\\
&\quad+\; \alpha^{2}\gamma(\omega_0{-}\omega_a)\,
   \mathcal{D}\!\big[\tilde{\sigma}_{-}\tilde{a}^{\dagger}\big]\tilde{\rho}(t)
 \;+\; \alpha^{2}\gamma(-\omega_0{+}\omega_a)\,
   \mathcal{D}\!\big[\tilde{\sigma}_{+}\tilde{a}\big]\tilde{\rho}(t)\nonumber\\
&\quad+\; \frac{\alpha^{4}}{4}\,\gamma(\omega_0{+}2\omega_a)\,
   \mathcal{D}\!\big[\tilde{\sigma}_{-}\tilde{a}^{2}\big]\tilde{\rho}(t)
 \;+\; \frac{\alpha^{4}}{4}\,\gamma(-\omega_0{-}2\omega_a)\,
   \mathcal{D}\!\big[\tilde{\sigma}_{+}\tilde{a}^{\dagger 2}\big]\tilde{\rho}(t)\nonumber\\
&\quad+\; \frac{\alpha^{4}}{4}\,\gamma(\omega_0{-}2\omega_a)\,
   \mathcal{D}\!\big[\tilde{\sigma}_{-}\tilde{a}^{\dagger 2}\big]\tilde{\rho}(t)
 \;+\; \frac{\alpha^{4}}{4}\,\gamma(-\omega_0{+}2\omega_a)\,
   \mathcal{D}\!\big[\tilde{\sigma}_{+}\tilde{a}^{2}\big]\tilde{\rho}(t).
\label{eq:fullmasterApp}
\end{align}
For a bosonic thermal bath with spectral density $J(\Omega)$ at temperature $T$, the dissipative spectrum entering the master equation is
\begin{align}
\gamma(\omega)=
\begin{cases}
J(|\omega|)\,\big[\bar{n}(|\omega|)+1\big], & \omega>0,\\[4pt]
J(|\omega|)\,\bar{n}(|\omega|), & \omega<0.
\end{cases}
\end{align}
Here $J(\omega)\ge 0$ is the bath spectral density, and the sign of $\omega$ distinguishes downward $(\omega)$ from upward $(-\omega)$ transitions. This definition obeys the KMS detailed-balance relation $\gamma(-\omega)=e^{-\omega/T}\,\gamma(\omega)$.
For a frequency-independent (flat) bath, $J(\omega)=\kappa$, where $\kappa$ is the coupling strength.

We now adiabatically eliminate the TLS in the dressed interaction picture obtained in Eq. (\ref{eq:fullmasterApp}). 
The carrier block at $\omega_{0}$ damps the TLS populations at the fast rate
\begin{equation}
\gamma_{0}=\gamma(\omega_{0})+\gamma(-\omega_{0}),
\label{A22}
\end{equation}
	\begin{figure}[!htbp]
		\begin{center}
			\leavevmode
			\includegraphics[width=0.48\textwidth,angle=0]{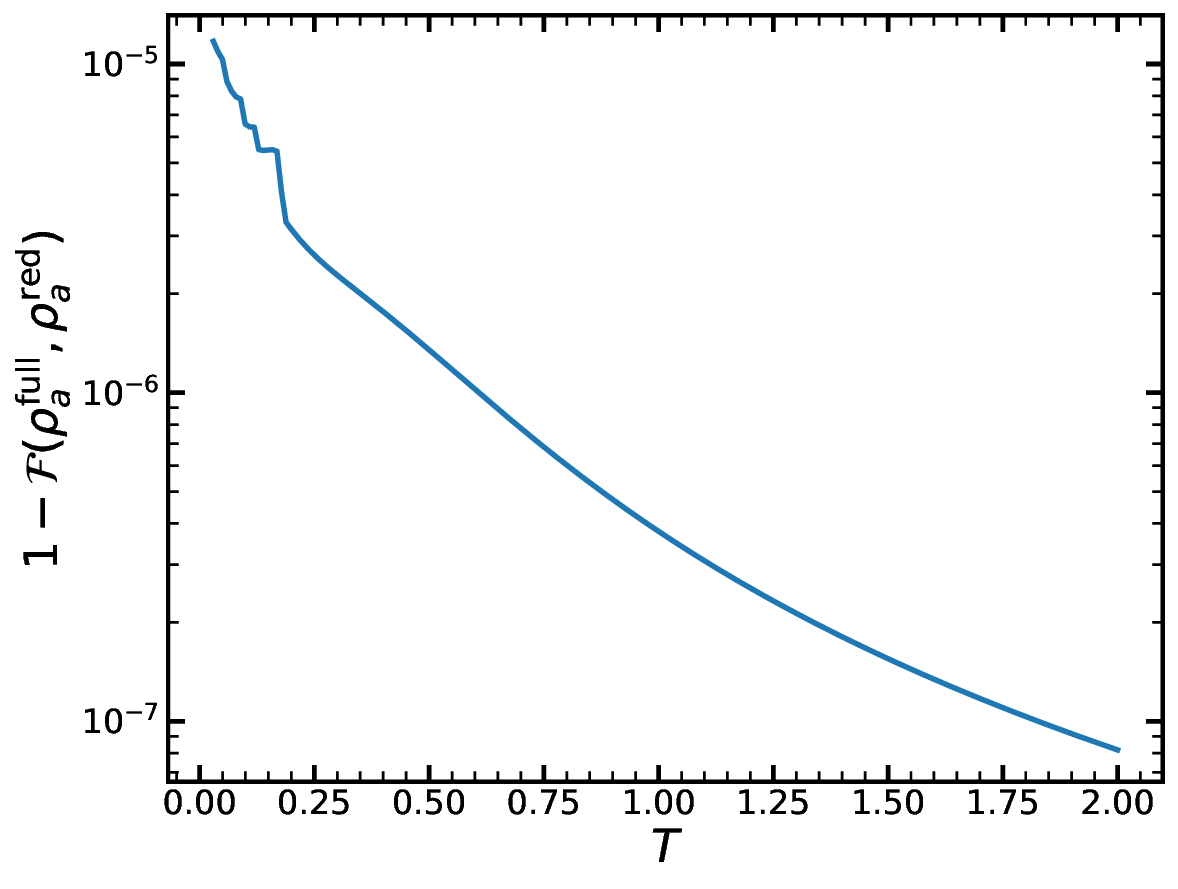}
			\caption{{\color{black}Log–scale infidelity $1-\mathcal{F}(\rho_a^{\mathrm{full}},\rho_a^{\mathrm{red}})$ between the steady–state oscillator obtained from the full dressed master equation~(\ref{eq:fullmasterApp}) and from the oscillator–only equation after adiabatic elimination~(\ref{eq:redmaster}), plotted versus bath temperature $T$.
            Both generators are written in the same dressed frame and Lamb shifts are neglected. A flat bath spectrum with detailed balance is used,
            $\gamma(\Delta\!>\!0)=\kappa\big(\bar n_T(\Delta)+1\big)$ and $\gamma(\Delta\!<\!0)=\kappa\,\bar n_T(|\Delta|)$ with $\bar n_T(\nu)=1/(\mathrm{e}^{\nu/T}-1)$. Representative parameters: $\omega_0=1$, $\omega_a=0.1$, $\alpha=g/\omega_a=0.1$, $\kappa=10^{-2}$, oscillator cutoff $N=50$. The infidelity remains $\lesssim 10^{-5}$ across the range shown, confirming the accuracy of the reduced model under the secular and time–scale separation assumptions.}}\label{fig:fig7}
		\end{center}
	\end{figure}
whereas the sideband and two–phonon blocks are slow, their norms being suppressed by $\alpha^{2}$ and $\alpha^{4}$, respectively. We therefore split the Liouvillian as $\mathcal{L}=\mathcal{L}_{0}+\mathcal{L}_{\mathrm{sb}}$, with
\begin{equation}
\mathcal{L}_{0}\tilde{\rho}
=\gamma(\omega_{0})\,\mathcal{D}\!\big[\big(1-\tfrac{\alpha^{2}}{2}-\alpha^{2}\tilde{a}^{\dagger}\tilde{a}\big)\,\tilde{\sigma}_{-}\big]\tilde{\rho}
+\gamma(-\omega_{0})\,\mathcal{D}\!\big[\big(1-\tfrac{\alpha^{2}}{2}-\alpha^{2}\tilde{a}^{\dagger}\tilde{a}\big)\,\tilde{\sigma}_{+}\big]\tilde{\rho},
\label{A23}
\end{equation}
and $\mathcal{L}_{\mathrm{sb}}$ equal to the sum of all one– and two–phonon dissipators at $\omega_{0}\pm\omega_{a}$ and $\omega_{0}\pm 2\omega_{a}$ given in Eq. (\ref{eq:fullmasterApp}).
Let $\tilde{\rho}(t)$ be the dressed joint state and introduce the projection onto the slow manifold
\begin{equation}
\mathcal{P}\tilde{\rho}=\operatorname{Tr}_{\mathrm{TLS}}(\tilde\rho)\otimes\tilde{\rho}^{\,\mathrm{ss}}_{\mathrm{TLS}},\qquad
\tilde{\rho}^{\,\mathrm{ss}}_{\mathrm{TLS}}=p_{e}\,|e\rangle\!\langle e|+p_{g}\,|g\rangle\!\langle g|,
\label{A24}
\end{equation}
with steady TLS populations set by the carrier rates
\begin{equation}
p_{e}=\frac{\gamma(-\omega_{0})}{\gamma(\omega_{0})+\gamma(-\omega_{0})},\qquad p_{g}=1-p_{e}.
\label{A25}
\end{equation}
Here $\tilde{\rho}^{\,\mathrm{ss}}_{\mathrm{TLS}}$ is the stationary state of the carrier block $\mathcal{L}_0$ at frequency $\omega_0$, so that $\mathcal{P}\mathcal{L}_0=0$; the oscillator part evolves only through the slower sideband and two-phonon blocks.
Writing $\mathcal{Q}=\mathbb{I}-\mathcal{P}$ and defining the small parameter
\begin{equation}
\eta:=\max\!\left\{\frac{\alpha^{2}\gamma(\omega_{0}\!\pm\!\omega_{a})}{\gamma_{0}},\ \frac{\alpha^{4}\gamma(\omega_{0}\!\pm\!2\omega_{a})}{\gamma_{0}}\right\}\ll1.
\label{A26}
\end{equation}
Standard adiabatic–elimination (Nakajima–Zwanzig) at leading nonvanishing order gives \cite{10.1143/PTP.20.948, 10.1063/1.1731409}
\begin{equation}
\frac{d}{dt}\,\mathcal{P}\tilde{\rho}(t)
=\mathcal{P}\mathcal{L}_{\mathrm{sb}}\mathcal{P}\tilde{\rho}(t)\;+\;O(\eta^{2}),
\label{A27}
\end{equation}
since $\mathcal{P}\mathcal{L}_{0}=0$ and the correction
$-\mathcal{P}\mathcal{L}_{\mathrm{sb}}\mathcal{Q}\mathcal{L}_{0}^{-1}\mathcal{Q}\mathcal{L}_{\mathrm{sb}}\mathcal{P}$ is $O(\eta^{2})$ and is neglected here. Equivalently, on time scales $\gg \gamma_{0}^{-1}$ the joint state factorizes as
\(
\tilde{\rho}(t)\simeq \tilde{\rho}_{a}(t)\otimes\tilde{\rho}^{\,\mathrm{ss}}_{\mathrm{TLS}}.
\)
Tracing each dissipator in $\mathcal{L}_{\mathrm{sb}}$ over the TLS is then straightforward. For any oscillator operator $X$ one uses
\begin{equation}
\operatorname{Tr}_{\mathrm{TLS}}\!\left[\mathcal{D}[\tilde{\sigma}_{-}X]\,(\tilde{\rho}_{a}\!\otimes\!\tilde{\rho}^{\,\mathrm{ss}}_{\mathrm{TLS}})\right]
= p_{e}\,\mathcal{D}[X]\,\tilde{\rho}_{a},\qquad
\operatorname{Tr}_{\mathrm{TLS}}\!\left[\mathcal{D}[\tilde{\sigma}_{+}X]\,(\tilde{\rho}_{a}\!\otimes\!\tilde{\rho}^{\,\mathrm{ss}}_{\mathrm{TLS}})\right]
= p_{g}\,\mathcal{D}[X]\,\tilde{\rho}_{a},
\label{A28}
\end{equation}
which follows directly from $\Pi_{e}=\tilde{\sigma}_{+}\tilde{\sigma}_{-}$ and $\tilde{\rho}^{\,\mathrm{ss}}_{\mathrm{TLS}}=p_{e}\Pi_{e}+p_{g}\Pi_{g}$.
Applying (\ref{A28}) to the sideband and two–phonon blocks yields the dressed oscillator–only master equation
\begin{equation}
\frac{d\tilde{\rho}_{a}}{dt}
=\tilde{\gamma}_-\,\mathcal{D}[\tilde{a}]\,\tilde{\rho}_{a}
+\tilde{\gamma}_+\,\mathcal{D}[\tilde{a}^{\dagger}]\,\tilde{\rho}_{a}
+\tilde{\Gamma}_-\,\mathcal{D}[\tilde{a}^{2}]\,\tilde{\rho}_{a}
+\tilde{\Gamma}_+\,\mathcal{D}[\tilde{a}^{\dagger 2}]\,\tilde{\rho}_{a}
+\gamma_{\phi}\,\mathcal{D}[\tilde{a}^{\dagger}\tilde{a}]\,\tilde{\rho}_{a},
\label{eq:redmaster}
\end{equation}
with coefficients:
\begin{align}
\tilde{\gamma}_\mp&=\alpha^{2}\!\Big( \langle \tilde{\sigma}_+\tilde{\sigma}_- \rangle \,\gamma(\omega_{0}{\pm}\omega_{a})+\langle \tilde{\sigma}_-\tilde{\sigma}_+ \rangle\,\gamma(-\omega_{0}{\pm}\omega_{a})\Big),\qquad
\tilde{\Gamma}_\mp =\frac{\alpha^{4}}{4}\!\Big(\langle \tilde{\sigma}_+\tilde{\sigma}_- \rangle\,\gamma(\omega_{0}{\pm}2\omega_{a})+\langle \tilde{\sigma}_-\tilde{\sigma}_+ \rangle\,\gamma(-\omega_{0}{\pm}2\omega_{a})\Big),\nonumber\\
\gamma_{\phi}&=\alpha^{4}\!\Big(\langle \tilde{\sigma}_+\tilde{\sigma}_- \rangle\,\gamma(\omega_{0})+\langle \tilde{\sigma}_-\tilde{\sigma}_+ \rangle\,\gamma(-\omega_{0})\Big).&&\label{A29c}
\end{align}
The $\tilde{\gamma}_\mp$ arises from the one–phonon sidebands at $\omega_{0}\pm\omega_{a}$; $\tilde{\Gamma}_\mp$  comes from the two–phonon harmonics at $\omega_{0}\pm2\omega_{a}$; and the last term in (\ref{A29c}) is a small number–dephasing. The elimination is valid under the time–scale separation and secular resolution already assumed, namely
\begin{equation}
\gamma_{0}\gg \alpha^{2}\gamma(\omega_{0}\pm\omega_{a}),\qquad
\gamma_{0}\gg \alpha^{4}\gamma(\omega_{0}\pm2\omega_{a}),
\label{A30}
\end{equation}
and $|\omega_{a}|$ large compared with the induced damping rates so that cross–frequency terms average out. To validate the adiabatic elimination, we compare the steady state of the oscillator obtained from the full dressed master equation~(\ref{eq:fullmasterApp}) with that from the reduced oscillator equation~(\ref{eq:redmaster}. Figure~\ref{fig:fig7} shows the steady–state infidelity \(1-\mathcal{F}(\rho_a^{\rm full},\rho_a^{\rm red})\) versus temperature \(T\) for a flat spectrum; the error remains \(\lesssim 10^{-5}\) across the range considered, confirming the quantitative accuracy of the reduced model under the secular and time–scale separation assumptions. 
}

	\bibliography{diode}
	
    \end{document}